\documentclass[aps, prb]{revtex4}

\usepackage{reuse_styles}

\begin{document}
\title{Monte Carlo study of the ordering of the weakly anisotropic Heisenberg spin glass in magnetic fields}
\author{Daisuke Imagawa and Hikaru Kawamura}
\affiliation{Department of Earth and Space Science, Faculty of Science,
Osaka University, Toyonaka 560-0043,
Japan}
\date{\today}

\begin{abstract}
The ordering of the three-dimensional Heisenberg
spin glass with the weak random anisotropy in magnetic fields
is studied by extensive equilibrium Monte Carlo simulations.
Both the spin and the chirality are monitored. 
We find strong numerical evidence that a
replica symmetry breaking transition occurs in
the chiral sector, which accompanies the simultaneous spin-glass order.
Despite the similarity in the global symmetry, 
the ordering behavior of the  
weakly anisotropic Heisenberg spin glass
differs significantly from that of the strongly anisotropic 
Ising spin glass.
The obtained phase diagram in the temperature - magnetic field plane  is
similar to the experimental
phase diagram.
Our results highlight the importance of the chirality in the spin-glass
ordering of the Heisenberg-like spin glass, and support
the spin-chirality decoupling-recoupling scenario of spin-glass
transitions. 
\end{abstract}

\maketitle

\section{Introduction}

Spin glasses (SGs) are random magnets 
where ferromagnetic and anti-ferromagnetic
exchange interactions co-exist and compete.\cite{SGrev}
Experimentally, it is now well established that SG magnets  in zero field
exhibit a thermodynamic second-order phase transition at a nonzero 
temperature into the thermodynamic SG phase. By contrast, 
whether SG magnets exhibit a thermodynamic phase transition
in applied magnetic fields has been a long-standing, yet unsolved issue.
This issue is closely
related to the fundamental question of whether 
the SG ordered state in zero field
accompanies an ergodicity breaking not directly related
to the global symmetry of the Hamiltonian, {\it i.e.\/}, the
replica symmetry breaking (RSB).

The experimental evidence of an in-field  transition of SG
remains to be obscure.
For the strongly anisotropic
Ising-like SG, Fe$_{0.5}$Mn$_{0.5}$TiO$_3$,
the non-existence of an in-field SG transition was reported in
Ref.\cite{Nordblad95}.
Meanwhile, many of
real SG materials are more or less Heisenberg-like rather than Ising-like 
in the sense that the magnetic anisotropy is considerably weaker than
the isotropic exchange interaction.\cite{SGrev}
Recent experiments on such
weakly anisotropic Heisenberg-like SGs
suggested the occurrence of an in-field 
SG transition,\cite{Campbell99,Campbell02} in apparent contrast 
to Ref.\cite{Nordblad95}.
Setting aside the question of
the strict nature of the apparent SG ``transition''
observed experimentally in applied fields, 
it has been known that the ``transition line'' 
between the paramagnetic and the SG phases
is similar to the one obtained for
the mean-field Sherrington-Kirkpatrick (SK) model: \cite{KotSom} Namely,
in the weak field regime, the in-field transition temperature, $T_g(H)$,
is rapidly suppressed with increasing the field intensity $H$, as
$H\propto |1-T_g(H)/T_g(0)|^{3/2}$ (de Almeida-Thouless (AT) line \cite{AT}),
while in the  high field regime, $T_g(H)$ stays rather robust against $H$,
behaving as $H\propto |1-T_g(H)/T_g(0)|^{1/2}$ (Gabay-Toulouse (GT) line 
\cite{GT}). However, 
the reason why the mean-field results have given such a good description
of the phase boundary, including the values of the critical exponents
describing the phase boundary, 
has remained to be a mystery.

On the theoretical side,
most of numerical studies on the finite-ranged SG models in three
dimensions (3D) have focused on the properties of
the Ising SG.\cite{SGrev}
Since no global symmetry exists in the Ising SG under magnetic fields,
an in-field transition, if any, should be a pure  RSB transition.
Unfortunately, numerical simulations on
the Ising SG 
have been unable to give a 
definitive answer concerning the existence of
a thermodynamic SG transition in magnetic 
fields.\cite{GrHe91,Singh91,Kawashima92,Badoni93,Marinari98,Parisi98,Picco98,Houdayer99,Krzakala01,Lamarcq03,Cruz03}

For the isotropic 3D Heisenberg SG in zero field,
it has been believed for years that the SG transition occurs
only at zero temperature, $i.e.\/$,  $T_\mathrm{SG}=0$.
\cite{Banavar,McMillan,OYS,Matsubara91,Yoshino} Since
applied magnetic fields make the SG transition even more unlikely, one
has expected no phase transition to occur in applied fields, and hence,
until quite recently, 
no extensive numerical simulation has ever been performed for the 3D Heisenberg
SG in magnetic fields.
Meanwhile, recent studies have revealed
that the Heisenberg SG
possesses an important physical ingredient absent in the Ising SG,
{\it i.e.\/}, the {\it chirality\/}.\cite{Kawamura92,Kawamura95,Kawamura98,HK00R,MatsuHuku,HK02,KawaIma01,ImaKawa02,ImaKawa04}
In the chirality scenario
of Ref.\cite{Kawamura92,Kawamura98}, 
in particular, the chirality is claimed
to be a hidden order parameter of the
SG transition of real Heisenberg-like SG magnets: In the fully isotropic
Heisenberg SG, the spin and the chirality, though they are coupled at short
length scales, are eventually decoupled at long
length scales, and
the system exhibits a chiral-glass transition at a finite temperature
without accompanying the
standard SG order. The chiral-glass transition corresponds to the spontaneous
breaking of the 
$Z_2$ spin-reflection
symmetry with preserving the $SO(3)$ spin-proper-rotation symmetry.
In the more realistic weakly anisotropic system,
the Heisenberg spin, decoupled from the chirality in the isotropic system,
is ``recoupled'' to the chirality
at long length scales via the random magnetic anisotropy.
The SG order of the weakly anisotropic Heisenberg SG is then dictated
at long length scales by the chirality ordering of the isotropic system.
Some numerical support of 
such a spin-chirality decoupling-recoupling
scenario was already reported in zero
field.\cite{Kawamura92,Kawamura95,Kawamura98,HK00R,MatsuHuku}  

By contrast, some other groups claimed that the chiral-glass
transition of the 3D
isotropic Heisenberg SG already accompanied the standard SG order, which means
that the standard SG order occurs at a finite temperature simultaneously
with the chirality.\cite{Matsubara00,Nakamura02,LeeYoung,Berthier}
Note that this is in contrast to the 
earlier belief in the community that the SG transition occurs only at 
$T=0$ in the 3D Heisenberg SG. Ref.\cite{HK02} maintains, however, that
the SG order occurs at a temperature lower than the
chiral-glass transition temperature, {\it i.e.\/}, 
$T_{{\rm CG}}>T_{{\rm SG}}\geq 0$, 
and the controversy remains. 
 
Recently, the present authors performed the first extensive MC simulation of
the 3D Heisenberg SG in magnetic
fields, and have observed that the chiral-glass transition,
essentially of the same character as in the zero-field one, 
occurs at a finite temperature even in magnetic fields.
\cite{KawaIma01,ImaKawa02}
The chiral-glass transition line in  the temperature-magnetic field
phase diagram turned out to have
a striking resemblance to the GT-line 
observed experimentally, 
although the nature of the transition is entirely different from the
mean-field GT-line. 
Note that the fully isotropic Heisenberg SG in  fields
possesses 
the global $Z_2\times SO(2)$ symmetry, the chiral $Z_2$ referring to
the global spin-reflection with respect to the plane containing the
magnetic-field axis, and the $SO(2)$ referring to
the global spin-rotation around the magnetic-field axis. 

In the more realistic case of the weakly anisotropic Heisenberg SG, 
by contrast, 
there no longer remains any global symmetry in fields. Hence, 
from  symmetry, the
situation is the same as that of the well-studied Ising SG. 
Meanwhile, in view of the fact that the Heisenberg SG possesses the 
nontrivial chiral degree of freedom which is totally absent in the Ising SG,
the question of whether the ordering properties of the
weakly anisotropic Heisenberg SG in fields
are essentially the same as those of the
Ising SG in fields 
seems not so trivial. This question is further promoted by the
apparently contradicting experimental observations on the Ising-like and
weakly anisotropic Heisenberg-like SGs.\cite{Nordblad95,Campbell99,Campbell02}

In the present paper,  we study both the spin-glass and the chiral-glass
orderings of the weakly anisotropic Heisenberg SG in magnetic
fields by extensive equilibrium
Monte Carlo (MC) simulations\cite{ImaKawa04}.
We  find a clear numerical
evidence that a finite-temperature RSB transition occurs
in the chiral sector, which also accompanies the simultaneous SG 
order. Thus, in spite of the similarity in the symmetry properties,
the ordering properties of the weakly anisotropic Heisenberg SG model 
turn out to be 
quite different from those of the standard Ising SG. This highlights
the importance of the chirality.

The paper is organized as follows. In \S 2, we introduce our model and explain
the details of the MC simulation.
Various physical quantities calculated in the simulation are defined in
\S 3, and the results
of our numerical simulation are presented in \S 4. In \S 5, we
perform the scaling analysis of the critical properties of the transition,
and construct a phase diagram of the model in the temperature - magnetic field
plane. Section 6  is devoted to
summary and discussion.

\section{The model and the method}
\label{Sec_model+method}

In this section, we introduce our model and explain some of
the details of our numerical method.
The model we consider is the isotropic classical Heisenberg
model on a 3D simple cubic lattice defined by the
Hamiltonian,
\begin{equation}
{\cal H}=
-\sum_{<ij>}(
J_{ij}\vec{S}_i\cdot\vec{S}_j
+ \sum_{\mu,\nu=x,y,z}D_{ij}^{\mu\nu}S_{i\mu}S_{j\nu})
- H\sum_{i=1}^N S_{iz}\ \ ,
\label{Hamiltonian}
\end{equation}
where $\vec{S}_i=(S_{ix},S_{iy},S_{iz})$ is a three-component unit vector,
and $H$ is the intensity of magnetic field applied along
the $z$ direction.
The isotropic nearest-neighbor exchange coupling
$J_{ij}$ is assumed to take either
the value $J$ or $-J$ with equal probability, 
while the  nearest-neighbor random exchange anisotropy
$D_{ij}^{\mu\nu}$'s ($\mu,\ \nu$=$x,y,z$ are spin-component indices)
are assumed to be
uniformly distributed in the range $[-D:D]$,
where $D$ is the typical intensity of the anisotropy.
We impose the relation
$D_{ij}^{\mu\nu}=D_{ji}^{\mu\nu}=D_{ij}^{\nu\mu}$.

We perform equilibrium MC simulations on this model.
In the present simulation, we fix $D/J=0.05$, which is a typical value of $D$
of real Heisenberg-like SG materials. 
Simulations are then performed for a variety of field intensities
in the range $H/J=0.02 - 3.0$.
The lattices studied are simple-cubic lattices with $N=L^3$ sites
with $L=4$, 6, 8, 10, 12 and 16 with periodic boundary conditions.
Sample average is taken over 64-800 independent bond realizations,
depending on the system size $L$ and the field intensity $H$.
Limited amount of data are also taken for $L=20$ in some cases
(32 samples) to check
the size dependence of physical quantities.

To facilitate efficient thermalization, we combine the standard
heat-bath method  with the temperature-exchange technique.\cite{TempExMC} The
temperature-exchange trial is performed every heat-bath sweep. Typically,
for the size $L=16$ ($L=20$), we discard initial $8\times 10^5$ 
($13\times 10^5$) heat-bath sweeps and the temperature-exchange trials
for equilibration, and use subsequent 
$8\times 10^5$ ($13\times 10^5$) heat-bath 
sweeps and the temperature-exchange trials in  calculating various 
physical quantities.
Care is taken to be sure that
the system is fully equilibrated.
Equilibration is checked by the following procedures:
First, we monitor the system to travel back and forth
many times during the
the temperature-exchange process (typically more than 10 times)
between the maximum and minimum temperature points, and at the same time check
that the relaxation
due to the standard heat-bath updating
is reasonably fast at the highest temperature,
whose relaxation time is of order $10^2$ Monte Carlo steps
per spin (MCS). This guarantees that  different parts of
the phase space are sampled in each ``cycle'' of the temperature-exchange
run. Second, we check
the stability of the results against at least three times longer runs
for a subset of samples. Error bars of
physical quantities are estimated by the sample-to-sample statistical
fluctuation over the bond realizations.
Further details of our MC simulations
are given in Table \ref{table-condition}.
%
%
%

\begin{table}
\caption{Details of our MC simulations.
$H/J$ represents the magnetic-field intensity, $L$ the lattice size,
$N_{\rm s}$ the total number of samples,  $N_{\rm T}$ the total number of
temperature points used in the temperature-exchange run,
$T_{\max}/J$ and $T_{\min}/J$  the maximum and minimum
temperatures in the temperature-exchange run.}
\label{table-condition}
\begin{center}
\begin{tabular}{|c|c|c|c|c|c|}
\hline
$H/J$ & $L$ & $N_{\rm s}$ & $N_{\rm T}$ & $T_{\max}/J$ & $T_{\min}/J$ \\
\hline
           &  4 & 800 & 26 & 0.475 & 0.085 \\
           &  6 & 800 & 26 & 0.475 & 0.085 \\
           &  8 & 600 & 26 & 0.475 & 0.113  \\
0.05       & 10 & 384 & 42 & 0.40 & 0.115 \\
           & 12 & 256 & 52 & 0.40 & 0.115 \\
           & 16 & 180 & 50 & 0.35 & 0.12   \\
           & 20 & 32 & 50 & 0.35 & 0.1775  \\
\hline
           &  4 & 400 & 26 & 0.475 & 0.085 \\
           &  6 & 400 & 26 & 0.475 & 0.085 \\
0.5        &  8 & 400 & 26 & 0.475 & 0.113  \\
           & 10 & 300 & 42 & 0.40 & 0.115 \\
           & 12 & 256 & 52 & 0.40 & 0.115 \\
           & 16 &  64 & 50 & 0.35 & 0.12   \\
           & 20 & 32 & 50 & 0.35 & 0.1775  \\
\hline
           &  4 & 400 & 26 & 0.475 & 0.085 \\
           &  6 & 400 & 26 & 0.475 & 0.085 \\
3.0        &  8 & 400 & 26 & 0.475 & 0.113  \\
           & 10 & 300 & 42 & 0.40 & 0.115 \\
           & 12 & 256 & 52 & 0.40 & 0.115 \\
           & 16 &  64 & 50 & 0.35 & 0.125  \\
\hline
\end{tabular}
\end{center}
\end{table}

\section{Physical Quantities}
\label{Sec_PhysQuant}

In this section, we
define various physical quantities calculated in our
simulations.

\subsection{Chirality-related quantities}

We begin with the definition of the chirality.
The local chirality at the $i$-th site and in the $\mu$-th
direction, $\chi_{i\mu}$, is defined for three neighboring Heisenberg 
spins by the scalar
\begin{equation}
\chi_{i\mu}=
\vec{S}_{i+\vec{e}_{\mu}}\cdot
(\vec{S}_i\times\vec{S}_{i-\vec{e}_{\mu}}),
\label{def_local_chiral}
\end{equation}
where $\vec{e}_{\mu}\ (\mu=x,y,z)$ denotes a unit vector along the
$\mu$-th axis. By this definition, there are in total $3N$ local
chiral variables in the system.
The local chirality amplitude is then defined by
\begin{equation}
\bar{\chi}=\sqrt{
\frac{1}{3N}\sum_{i=1}^N \sum_{\mu=x,y,z}[\thermav{\chi_{i\mu}^2}]\ \ ,
}
\label{def_locx}
\end{equation}
where $\thermav{\cdots }$ represents the thermal average and $[\cdots ]$ 
represents the average over the bond disorder.
The local chirality amplitude gives us the information of 
the extent of the non-coplanarity
of local spin structures.

By considering two independent systems (``replicas'') described by
the same Hamiltonian,
one can define an overlap of the chiral variable
via the relation,
\begin{equation}
q_{\chi}=
\frac{1}{3N}\sum_{i=1}^N\sum_{\mu=x,y,z}\chi_{i\mu}^{(a)}\chi_{i\mu}^{(b)}\ \ ,
\label{def_q_chi}
\end{equation}
where $\chi_{i\mu}^{(a)}$ and  $\chi_{i\mu}^{(b)}$ represent the chiral
variables of the replicas ``a'' and ``b'', respectively.
In our simulations, we prepare the two replicas a and b by
running two independent sequences of  systems
in parallel with different spin initial conditions and
different sequences of random numbers.

Since the present model does not possess any global symmetry, an odd quantity
$[\thermav{q_\chi}]$ is generally non-zero even in the hight-temperature phase.
Taking this effect into consideration,
the chiral-glass order parameter may be defined by
\begin{equation}
\tilde{q}_{\chi}^{(2)}=[\thermav{(q_{\chi}-[\thermav{q_{\chi}}])^2}]\ \ .
\end{equation}
The associated chiral-glass susceptibility, normalized by the local
amplitude $\bar \chi$, is defined by
\begin{equation}
\tilde \chi_\chi=3N\frac{\tilde{q}_\chi^{(2)}}{\bar{\chi}^4}\ \ .
\label{def_xcg}
\end{equation}
The Binder ratio of the chirality is defined by
\begin{equation}
g^\prime_{\chi}=
\frac{1}{2}
\left(3-\frac{\tilde{q}_\chi^{(4)}}
{(\tilde{q}_\chi^{(2)})^2}\right)\ \ ,
\label{def_g_chi2}
\end{equation}
where
\begin{equation}
\tilde{q}_\chi^{(4)}=[\thermav{(q_\chi-[\thermav{q_\chi}])^4}]\ \ .
\end{equation}
Here, $g^\prime_\chi$
is normalized so that, in the thermodynamic limit, 
it vanishes in the high-temperature phase and gives unity in the
ordered phase if the ordered state is non-degenerate.
The distribution function of the chiral overlap
$q_{\chi}$ is defined by
\begin{equation}
P_\chi(q^{\prime}_{\chi})=[\langle\delta(q_\chi^{\prime}-q_{\chi})\rangle]\ \ .
\label{def_Pq_chi}
\end{equation}

We consider the Fourier-transformed 
two-point chiral-glass correlation function
${\cal C}_\mathrm{CG}^{\mu\nu}(\vec{k})$ 
between the two local chiral variables in the $\mu$-th and in the $\nu$-th 
directions. While one can define various types of correlation functions 
depending on the relative directions of the chiral variables ($\mu , \nu)$ 
and the
direction of $\vec k$, we consider here the parallel component 
${\cal C}_\mathrm{CG}^{\parallel}(\vec{k})$ where
$\mu$ and  $\nu$ are both parallel with $\vec{k}$.
Here, we put $\vec{k}$ parallel with the $x$-direction, $\vec k=(k,0,0)$, 
so that $\mu=\nu=x$.
Then, ${\cal C}_\mathrm{CG}^{\parallel}(\vec{k})$ can be written in terms of
the $k$-dependent chiral overlap $q_\chi(\vec k)$, 
\begin{equation}
{\cal C}_\mathrm{CG}^{\parallel}(\vec{k})=[<|q_\chi(\vec k)|^2>],
\end{equation}
\begin{equation}
q_\chi(\vec k)=\frac{1}{N}\sum_{i=1}^N\chi_{ix}^{(a)}\chi_{ix}^{(b)}
\exp (i\vec k\cdot \vec r_i),
\end{equation}
where $\vec{r}_i=(x_i,y_i,z_i)$ denotes the
position vector of the chiral variable at the $i$-th site.
The associated chiral correlation length, $\xi_\chi$,
is defined by,
\begin{equation}
\xi_\chi=
\frac{1}{2\sin(k_\mathrm{m}/2)}
\sqrt{\frac{{\cal C}_\mathrm{CG}^\parallel(\bf{0})}
{{\cal C}_\mathrm{CG}^\parallel(\vec{k}_\mathrm{m})}-1}
\label{def_xi_xp}
\ \ ,
\end{equation}
where $\vec{k}_\mathrm{m}=(2\pi/L, 0, 0)$ is a wavevector
of the minimum magnitude.

  In order to study the equilibrium dynamics of the model,
we compute the equilibrium autocorrelation function of the chirality 
defined by
\begin{eqnarray}
\tilde{C}'_\chi(t)
&=&
\left[\left\langle \frac{1}{3N}\sum_{i=1}^N\sum_{\mu=x,y,z}
\chi_{i\mu}(t_0)\chi_{i\mu}(t+t_0)\right\rangle\right] \ - \ 
[\thermav{q_\chi}]\ \ ,
\label{def_time_Cx}
\end{eqnarray}
where the ``time'' $t$ is measured in units of MCS.
This chiral autocorrelation function is an 
``odd'' quantity not invariant under
the global flipping of the chirality. 
The corresponding ``even'' time correlation functions 
which is invariant
under the global flipping of the chirality, may be defined by 
\begin{eqnarray}
\tilde{q}_\chi^{(2)'}(t)
&=&
\left[\left\langle\left(\frac{1}{3N}\sum_{i=1}^N\sum_{\mu=x,y,z}
\chi_{i\mu}(t_0)\chi_{i\mu}(t+t_0)\right)^2\right\rangle\right] 
\ - \ [\thermav{q_\chi^2}]\ \ .
\label{def_time_qx2}
\end{eqnarray}
In displaying the data, 
we normalize these time correlation functions by their values 
at a unit time $t=1$, {\it i.e.\/}, we put
\begin{eqnarray}
C_\chi(t)=C'_\chi(t)/C'_\chi(1), \ \ \ \ \ \ 
\tilde{q}_\chi^{(2)}(t)=\tilde{q}_\chi^{(2)'}(t)/\tilde{q}_\chi^{(2)'}(1).
\end{eqnarray}

Note that in the above definitions of the time-correlation functions (13)
and (14),
the second terms, $[\thermav{q_\chi}]$
and $[\thermav{q_{\chi}^2}]$, have been subtracted,
which are nonzero even in the high-temperature phase in the
$L\rightarrow \infty$ limit due to the absence of any global symmetry.
This subtraction
guarantees that both $\tilde{C}_\chi(t)$ and 
$\tilde q^{(2)}_\chi(t)$ decay to zero as
$t\rightarrow \infty$  in the
high-temperature phase. In the possible ordered phase,
by contrast,
both $\tilde{C}_\chi(t)$ and  
$\tilde q^{(2)}_\chi(t)$ decay to zero {\it if\/} 
the ordered
state does not
accompany the RSB, but tend to finite positive values {\it if\/} the
ordered state accompanies the RSB.
The latter property arises because,  in the presence of RSB, 
the $t\rightarrow \infty$ limits of the first terms of Eqs.(13) and (14)
are generally greater than the second terms, 
$[\thermav{q_\chi}]$ and $[\thermav{q_\chi^2}]$.

In computing the first terms of Eqs.(13) and (14), the simulation is performed
according to the standard heat-bath updating without the
temperature-exchange procedure, while
the starting spin configuration at $t=t_0$ is taken from
the equilibrium spin configurations
generated in our temperature-exchange MC runs.
The second terms of Eqs.(13) and (14)
are evaluated from the temperature-exchange MC runs.

\subsection{Spin-related quantities}

As in the case of the chirality, it is convenient to define an overlap
variable  for the
Heisenberg spin. In this case, the overlap
might naturally be defined
as a {\it tensor\/} variable $q_{\mu\nu}$
between the $\mu$ and $\nu$
components  ($\mu$, $\nu$=$x,y,z$) of the Heisenberg spin,
\begin{equation}
q_{\mu\nu}=\frac{1}{N}\sum_{i=1}^N  S_{i\mu}^{(a)}S_{i\nu}^{(b)}\ \ ,
\ \ (\mu=x,y,z),
\label{def_qmn}
\end{equation}
where $\vec{S}_i^{(a)}$ and $\vec{S}_i^{(b)}$ are the $i$-th
Heisenberg spins of the replicas a and b, respectively.

In terms of these tensor overlaps, the ``longitudinal'' (parallel to the
applied field) and the ``transverse'' (perpendicular to the applied field)
SG order parameters may be defined by
\begin{equation}
\tilde{q}_\mathrm{L}^{(2)} = [\thermav{(q_\mathrm{L}-[\thermav{q_\mathrm{L}}])^2}],\ \ \ \
q_\mathrm{L} = q_{zz}\ \ ,
\end{equation}
\begin{equation}
\tilde{q}_\mathrm{T}^{(2)} = [\thermav{(q_\mathrm{T}-[\thermav{q_\mathrm{T}}])^2}],\ \ \ \
q_\mathrm{T} = \sum_{\mu = x, y}q_{\mu\mu} = q_{xx} + q_{yy}\ \ .
\end{equation}
Note that as in the case of the chirality,
the expectation value of the first moment has been subtracted.
We also consider the
spin-glass susceptibility
for 
the transverse spin component defined by,
%
%
%
\begin{equation}
\tilde{\chi}_\mathrm{T}=N\tilde{q}^{(2)}_\mathrm{T}\ \ .
\label{def_XsgT}
\end{equation}
The longitudinal and the transverse Binder ratios are defined, respectively,  
by
\begin{eqnarray}
g^\prime_\mathrm{L}
&=&
\frac{1}{2}
\left(3 - \frac{\tilde{q}_\mathrm{L}^{(4)}}{(\tilde{q}_\mathrm{L}^{(2)})^2}\right)\ \ ,
\label{def_gL2}
\\
g^\prime_\mathrm{T}
&=&
\frac{1}{2}
\left(3 - \frac{\tilde{q}_\mathrm{T}^{(4)}}{(\tilde{q}_\mathrm{T}^{(2)})^2}\right)\ \ ,
\label{def_gT2}
\end{eqnarray}
where
\begin{eqnarray}
\tilde{q}_\mathrm{L}^{(4)}
&=&
[\thermav{(q_\mathrm{L}-[\thermav{q_\mathrm{L}}])^4}] \ \ ,\\
\tilde{q}_\mathrm{T}^{(4)}
&=&
[\thermav{(q_\mathrm{T}-[\thermav{q_\mathrm{T}}])^4}]\ \  .
\end{eqnarray}
Here, $g^\prime_\mathrm{L}$ and $g^\prime_\mathrm{T}$
are normalized so that,  in the thermodynamic limit,
they vanish in the high-temperature phase and give unity in the ordered state
if the ordered state is non-degenerate.

%
%

The full spin-overlap distribution function may be defined in the tensor
space with $3\times 3=9$ components. Here, we consider the 
spin-overlap distribution 
function for the longitudinal and the transverse components, each 
defined by
\begin{equation}
P_\mathrm{s}(q_\mathrm{L}^\prime)=[\thermav{\delta (q_\mathrm{L}^\prime-q_\mathrm{L})}]\ \ ,
\end{equation}
\begin{equation}
P_\mathrm{s}(q_\mathrm{T}^\prime)=[\thermav{\delta (q_\mathrm{T}^\prime-q_\mathrm{T})}]\ \ ,
\end{equation}
where $q_{\rm L}$ and $q_{\rm T}$ are defined by Eqs.(17) and (18),
respectively. 
In the possible SG ordered state of the 
{\it isotropic\/} system, $P_{\rm s}(q_{\rm T})$ develops a non-trivial
shape in the thermodynamic limit
due to the fact that the transverse-spin-overlap $q_{\rm T}$ transforms
non-trivially under the global spin rotation: See Refs. \cite{KawaLi01} 
and \cite{ImaKawa03} for details. 
But here, the system is anisotropic so that no
non-trivial structure arising from the uniform global spin-rotation 
is expected to arise in $P_{\rm s}(q_{\rm T})$ in the thermodynamic limit.

We consider the Fourier-transformed 
spin-glass correlation functions both for the longitudinal and the transverse
components,
${\cal C}_\mathrm{L}(\vec{k})$ and ${\cal C}_\mathrm{T}(\vec{k})$, 
which can be written in terms of
the $k$-dependent longitudinal and transverse spin-overlaps, 
$q_\mathrm{L}(\vec k)$ and $q_\mathrm{T}(\vec k)$, as, 
\begin{equation}
{\cal C}_\mathrm{L}(\vec{k})=[<|q_\mathrm{L}(\vec k)|^2>],\ \ \ \ \ 
{\cal C}_\mathrm{T}(\vec{k})=[<|q_\mathrm{T}(\vec k)|^2>],
\end{equation}
with
\begin{equation}
q_\mathrm{L}(\vec k)=\frac{1}{N}\sum_{i=1}^NS_{iz}^{(a)}S_{iz}^{(b)}
\exp (i\vec k\cdot \vec r_i),\ \ \ \ \ \ 
q_\mathrm{T}(\vec k)=\frac{1}{N}\sum_{i=1}^N 
\vec S_{i{\rm T}}^{(a)}\cdot \vec S_{i{\rm T}}^{(b)}
 \exp (i\vec k\cdot \vec r_i).
\end{equation}
where $\vec S_{i{\rm T}}=(S_{ix},S_{iy})$ represents the transverse ($xy$) 
component of the Heisenberg spin.
The associated longitudinal and the transverse spin correlation lengths,
$\xi_\mathrm{L}$ and $\xi_\mathrm{T}$,
are defined by
\begin{equation}
\xi_\mathrm{L}=
\frac{1}{2\sin(k_\mathrm{m}/2)}
\sqrt{\frac{{\cal C}_\mathrm{L}(\bf{0})}
{{\cal C}_\mathrm{L}(\vec{k}_\mathrm{m})}-1}\ \ ,
\label{def_xi_L}
\end{equation}
\begin{equation}
\xi_\mathrm{T}=
\frac{1}{2\sin(k_\mathrm{m}/2)}
\sqrt{\frac{{\cal C}_\mathrm{T}(\bf{0})}
{{\cal C}_\mathrm{T}(\vec{k}_\mathrm{m})}-1}\ \ ,
\label{def_xi_T}
\end{equation}
respectively.

The spin autocorrelation functions are defined
both for the longitudinal and the transverse components by
\begin{eqnarray}
\tilde{C}_\mathrm{L}(t)=\tilde{C}_\mathrm{L}'(t)/\tilde{C}_\mathrm{L}'(1),
\ \ \ \ 
\tilde{C}_\mathrm{L}'(t)
&=&
\frac{1}{N}\sum_{i=1}^N
[\thermav{S_{iz}(t_0)S_{iz}(t+t_0)}] - [\thermav{q_\mathrm{L}}]\ \ ,
\label{def_time_CL}
\\
\tilde{C}_\mathrm{T}(t)=\tilde{C}_\mathrm{T}'(t)/\tilde{C}_\mathrm{T}'(1), 
\ \ \ \ 
\tilde{C}_\mathrm{T}'(t)
&=&
\frac{1}{N}\sum_{i=1}^N
[\thermav{\vec{S}_{i{\rm T}}(t_0)\cdot\vec{S}_{i{\rm T}}(t+t_0)}] - 
[\thermav{q_{\rm T}}]\ \ .
\label{def_time_CT}
\end{eqnarray}
The second terms,  $[\thermav{q_\mathrm{L}}]$ and 
$[\thermav{q_{\rm T}}]$, 
have been subtracted in the same context
as in the definition of $\tilde{C}_\chi(t)$ 
for the chirality.
These spin autocorrelation functions are computed in the same way as
the chiral autocorrelation functions, and
are  normalized at their values at a unit time $t=1$.
The corresponding 
``even'' time correlation functions 
can also be defined,
though we skip their definitions here.

%
%
%
%
%
%


\section{Numerical results}
\label{Sec_Results}

In this section, we show the results of our MC simulations on the anisotropic
Heisenberg SG with $D/J=0.05$.

\subsection{Time correlation functions of the chirality}

\begin{figure}[ht]
  \leavevmode
  \begin{center}
    \begin{tabular}{l}
      \includegraphics[scale=0.7]{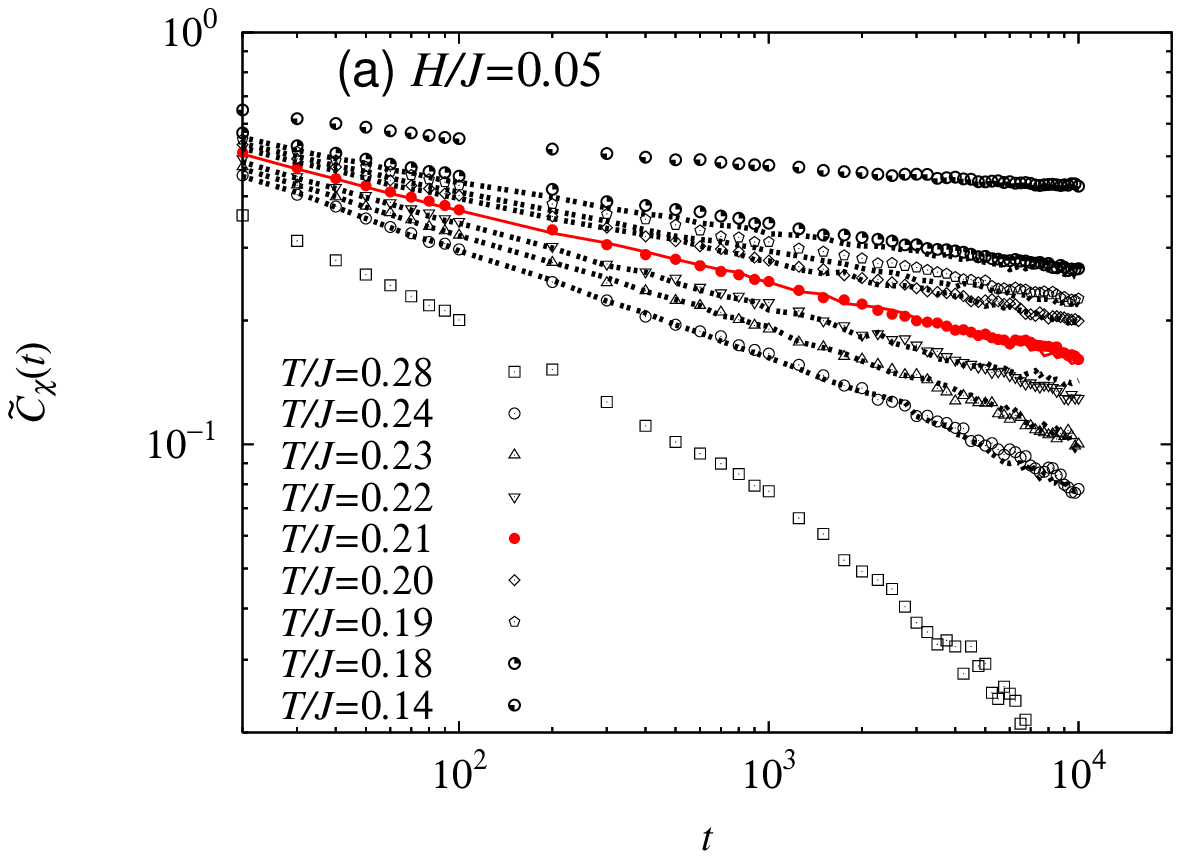}
    \end{tabular}
    \begin{tabular}{ll}
      \includegraphics[scale=0.7]{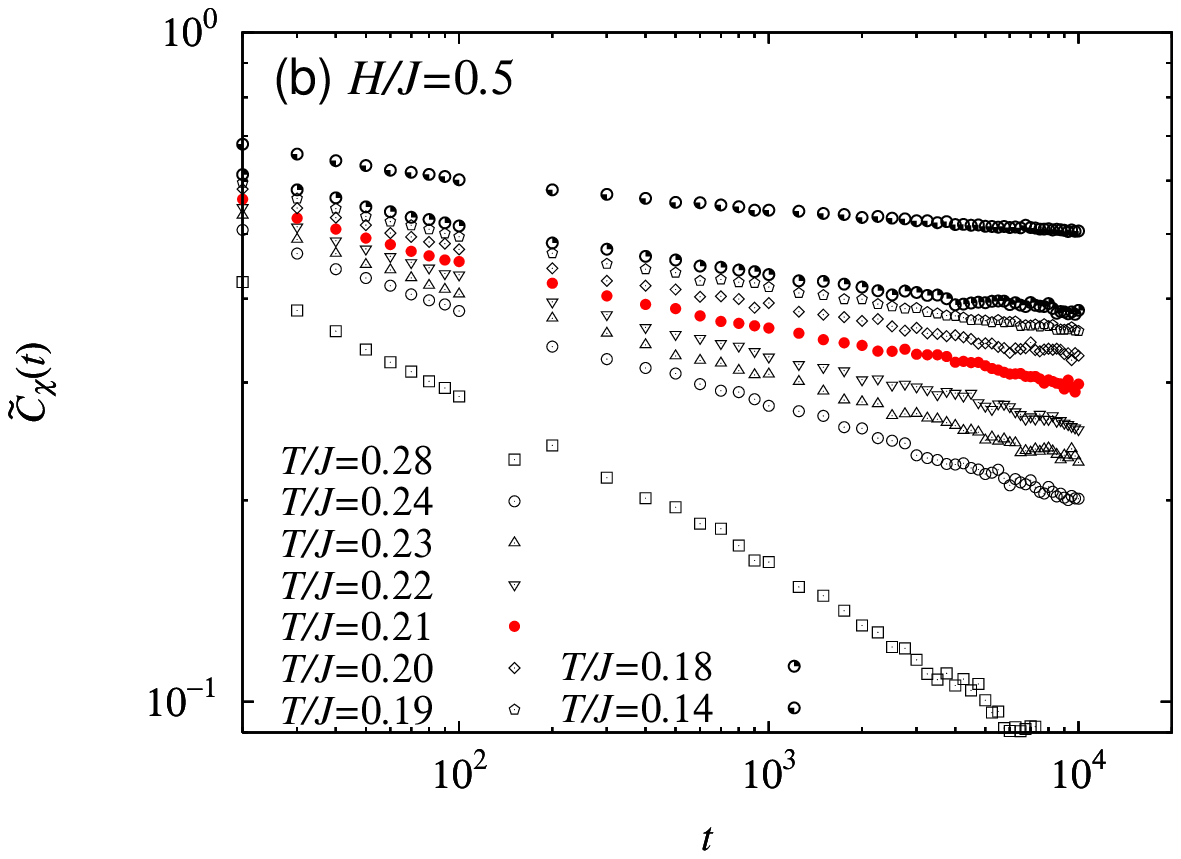} &
      \includegraphics[scale=0.7]{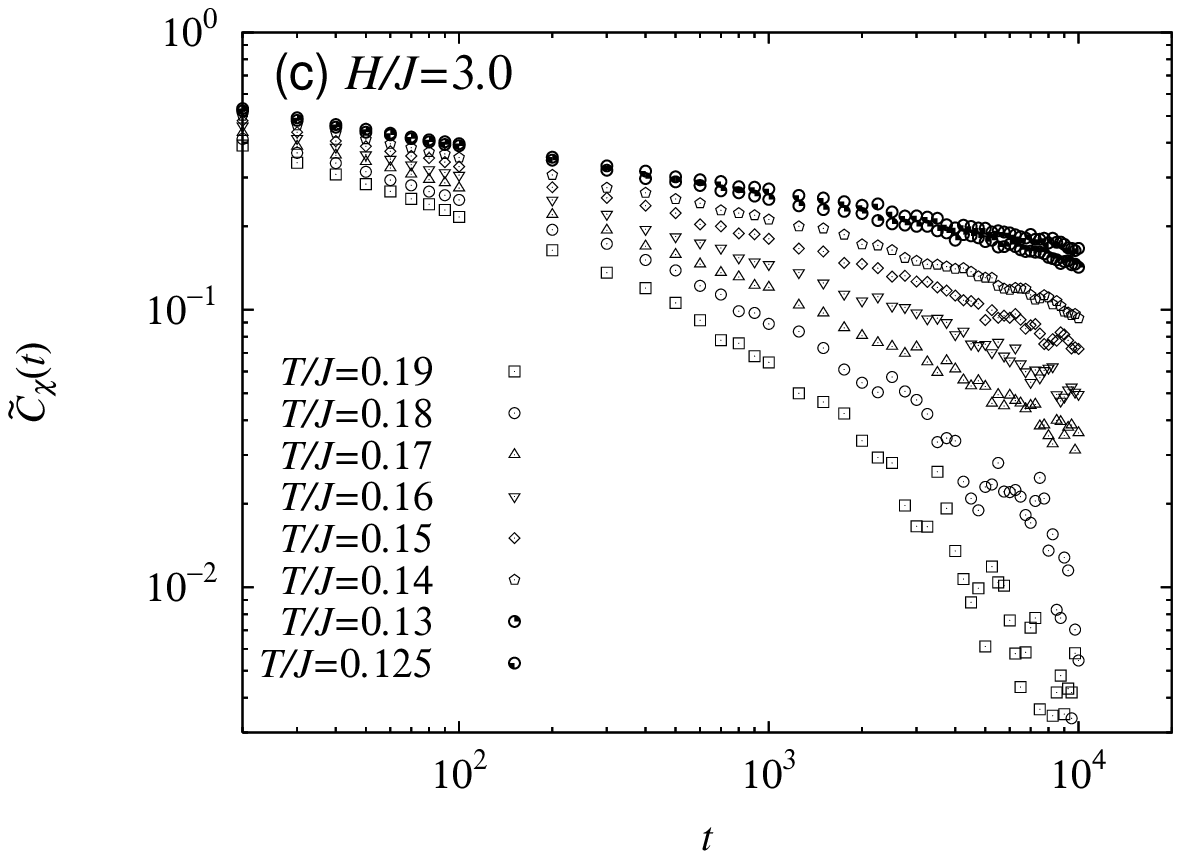}
    \end{tabular}
\caption{(Color online) 
Temporal decay of the autocorrelation
function of the chirality $\tilde{C}_{\chi}(t)$ defined by Eqs. (13)
and (15),
for the fields (a) $H/J=0.05$, (b) $H/J=0.5$ and (c) $H/J=3.0$.
The lattice size is $L=16$. The data at $T=T_g$
are given in red (by filled symbols).
In Fig.(a),
in order to check the finite-size effect,
the data of $L=20$ are plotted with lines
at temperatures $T/J=0.18$-$0.24$ with an interval of 0.01.
The estimated transition
temperature is $T_g/J\simeq 0.21$ both for $H/J=0.05$
and $H/J=0.5$.
}
    \label{fig_time_Cx_D005}
  \end{center}
\end{figure}

In Fig.\ref{fig_time_Cx_D005}, we show
the MC time dependence of the chiral autocorrelation function
$\tilde{C}_{\chi}(t)$ for the size $L=16$
on log-log plots, for the fields (a) $H/J=0.05$, (b) $H/J=0.5$
and (c) $H/J=3.0$, respectively.
In the case $H/J=0.05$, we have checked  that, in the time range shown, 
the data can be regarded as those of the bulk,
since no appreciable size effect is discernible between the data
of $L=16$ and $L=20$, the latter data being shown with lines in the figure. 
In Fig.2, we show  the MC time dependence of the corresponding even quantities,
$\tilde{q}_{\chi}^{(2)}(t)$, 
for the fields (a) $H/J=0.05$, (b) $H/J=0.5$
and (c) $H/J=3.0$, respectively. As can clearly be seen from these figures, 
in the investigated time range, the observed 
behavior of $\tilde{C}_{\chi}(t)$ is essentially the same as that of
$\tilde{q}_{\chi}^{(2)}(t)$. 

For the field $H/J=0.05$, as can be seen from the Figs. 1(a) and 2(a), 
$\tilde{C}_{\chi}(t)$ and $\tilde{q}_{\chi}^{(2)}(t)$
exhibit either a
down-bending or an up-bending behavior depending on whether the temperature is
higher or lower than a borderline value $T/J\simeq 0.21$.
while just at this borderline temperature
a straight-line behavior corresponding to a power-law decay is observed.
This indicates that the chirality exhibits a phase
transition into the low-temperature ordered phase where the
replica symmetry is spontaneously broken.
In the case of $H/J=0.5$, as can be seen from  Figs.1(b) and 2(b), 
the data at $T/J\lsim 0.24$ show a
slight up-bending tendency at a short time $t\simeq 10^2$ while the data at 
$T/J\gsim 0.21$ show a gradual down-bending tendency 
at longer times $t\simeq 10^3$, and 
the chiral transition temperature appears to lie somewhere between
$T/J\simeq 0.21$ and 0.24. The transverse spin autocorrelation, 
which is 
to be shown in Fig.3(c) below, however, suggests the transition temperature
$T_g\simeq 0.20-0.21$. Then, 
we finally estimate the transition temperature
$T_g/J= 0.21\pm 0.02$ for $H/J=0.5$ (see below). 
In the case $H/J=3.0$, by contrast, $\tilde{C}_{\chi}(t)$ and 
$\tilde{q}_{\chi}^{(2)}(t)$ always exhibits a down-bending behavior
in the temperature range studied,
suggesting the absence of a phase transition, at least 
in the temperature range $T/J\geq 0.125$.

Since very much similar behaviors are observed in $\tilde{C}_{\chi}(t)$ and
in $\tilde{q}_{\chi}^{(2)}(t)$,
we shall 
show in the following subsections the data of the autocorrelation function
$\tilde C(t)$ only.

\begin{figure}[ht]
  \leavevmode
  \begin{center}
    \begin{tabular}{l}
      \includegraphics[scale=0.7]{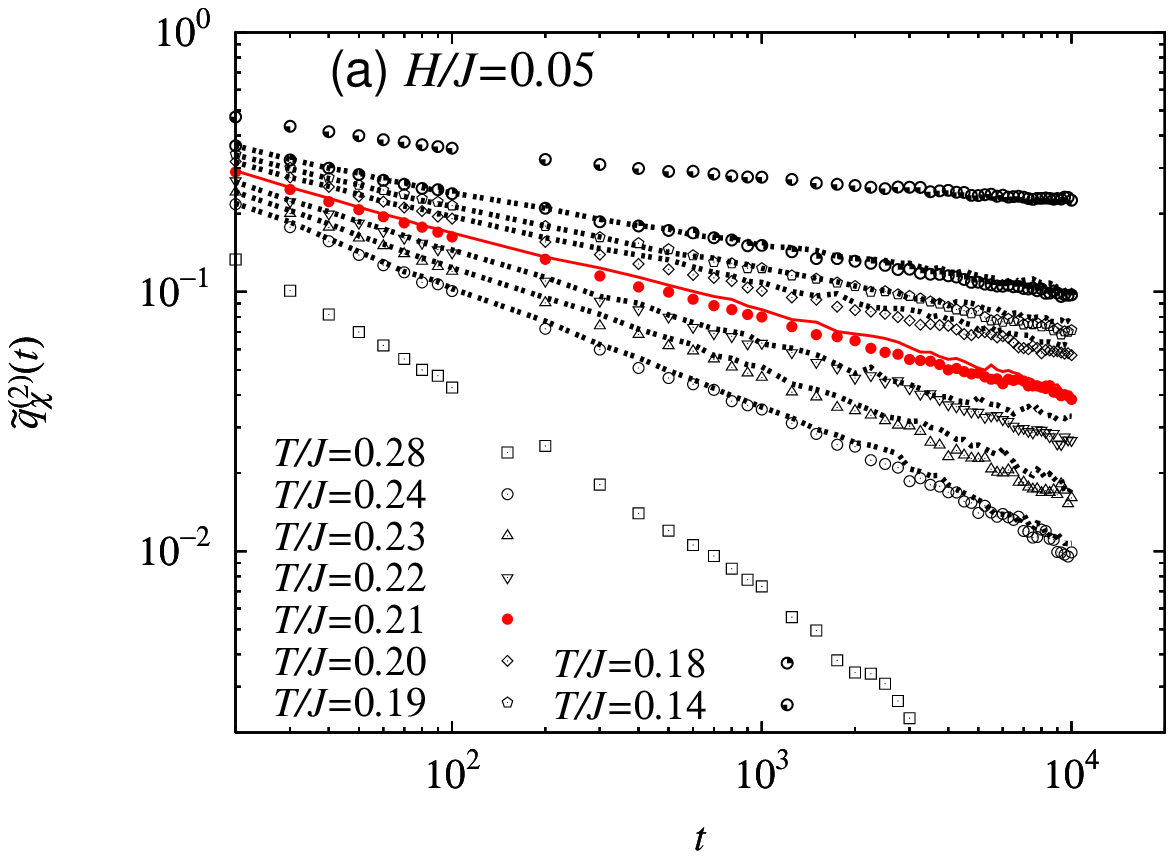}
    \end{tabular}
    \begin{tabular}{ll}
      \includegraphics[scale=0.7]{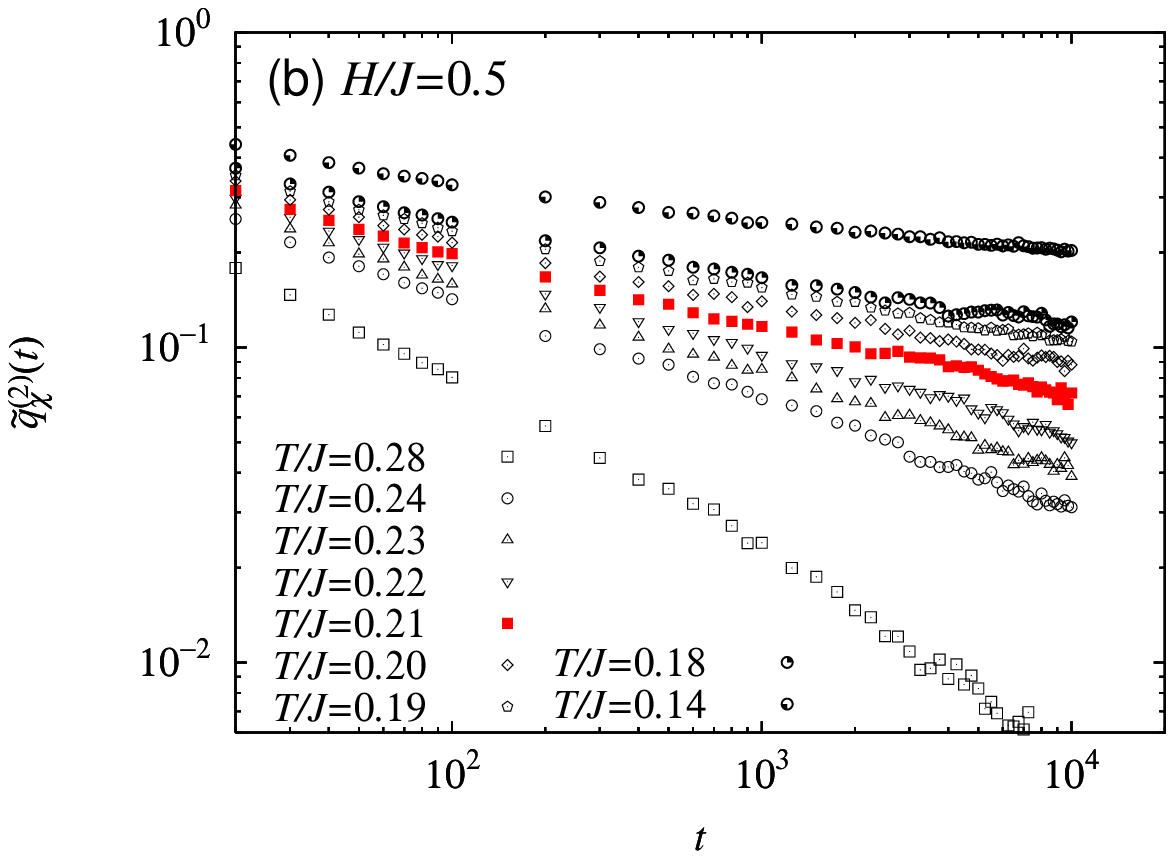} &
      \includegraphics[scale=0.7]{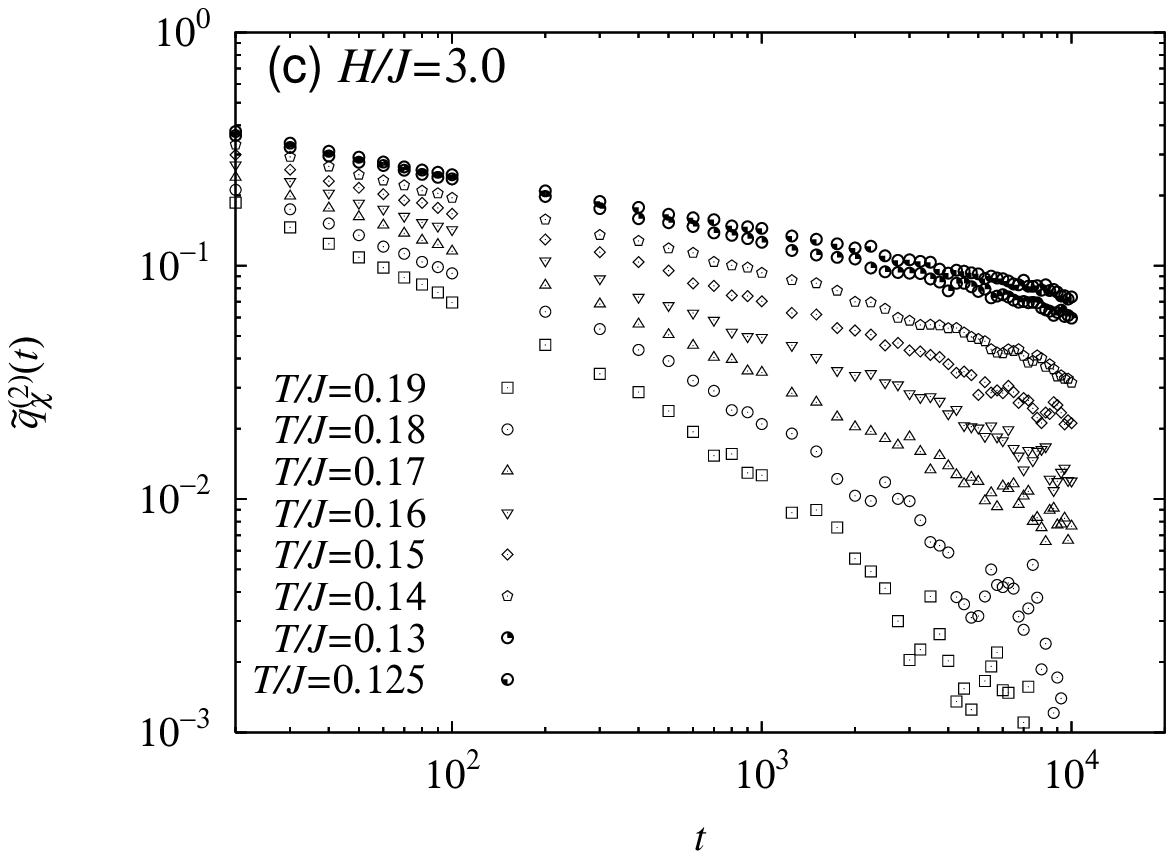}
    \end{tabular}
\caption{(Color online) 
Temporal decay of the time-correlation
function of the chirality $\tilde{q}_{\chi}^{(2)}(t)$ 
defined by Eqs.(14) and (15),
for the fields (a) $H/J=0.05$, (b) $H/J=0.5$ and (c) $H/J=3.0$.
The lattice size is $L=16$. The data at $T=T_g$
are given in red (by filled symbols). In Fig.(a),
in order to check the finite-size effect,
the data of $L=20$ are plotted with lines
at temperatures $T/J=0.18$-$0.24$ with an interval of 0.01.
The estimated transition
temperature is $T_g/J\simeq 0.21$ both for $H/J=0.05$
and $H/J=0.5$.
}
  \end{center}
\end{figure}

\subsection{Time correlation functions of the spin}

\begin{figure}[ht]
  \leavevmode
  \begin{center}
    \begin{tabular}{ll}
      \includegraphics[scale=0.7]{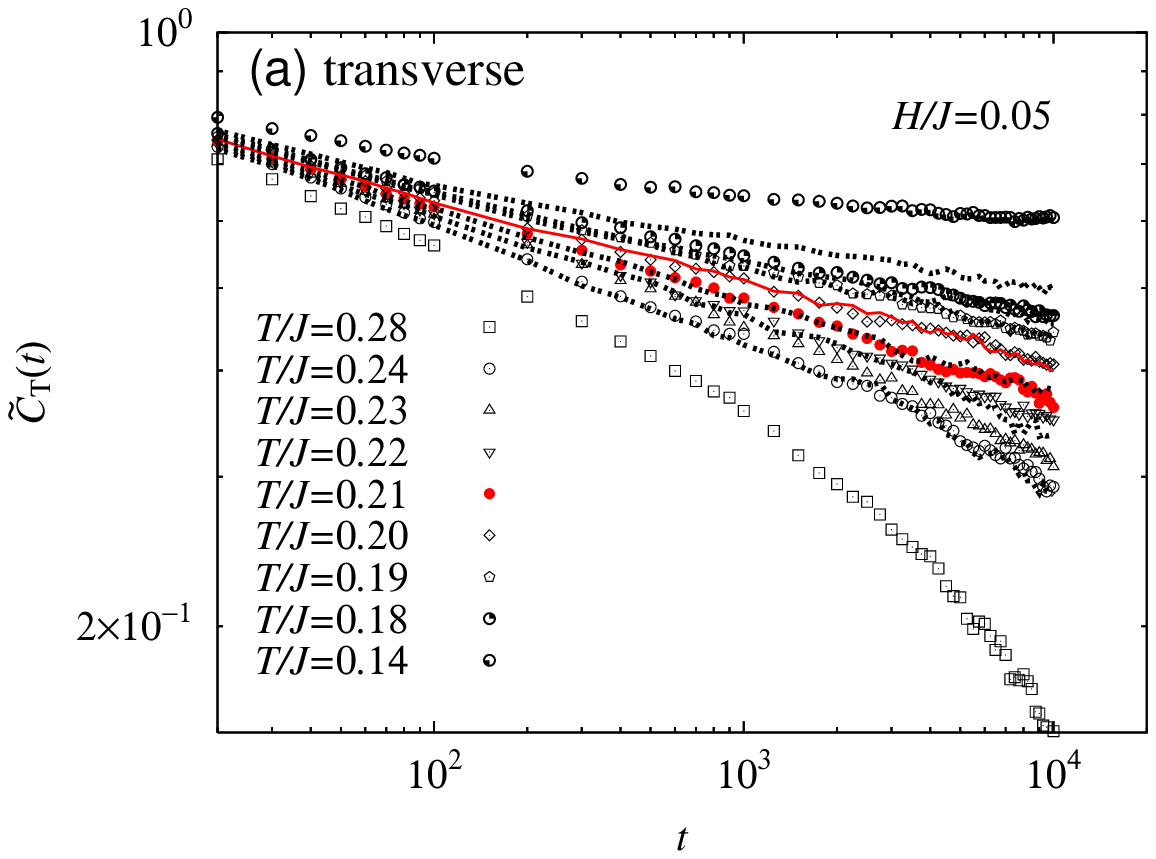} &
      \includegraphics[scale=0.7]{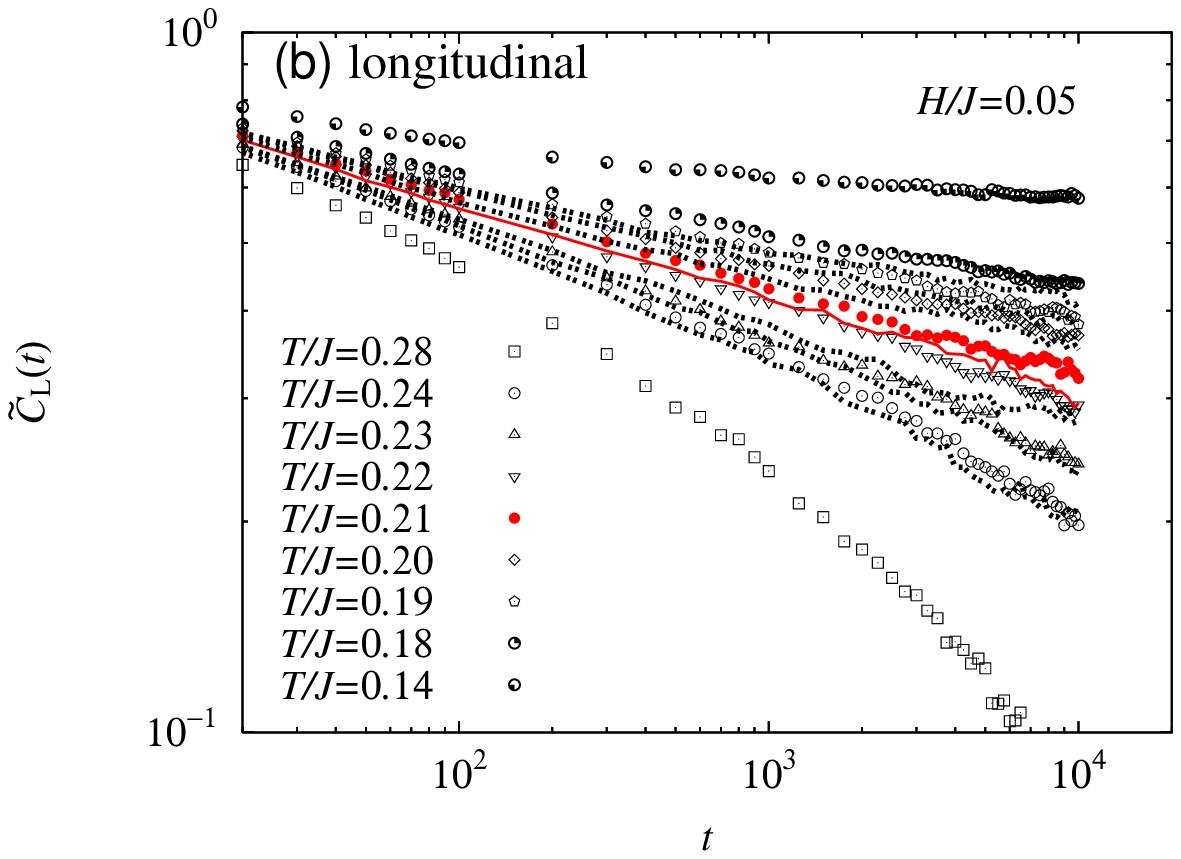} 
    \end{tabular}
    \begin{tabular}{ll}
      \includegraphics[scale=0.7]{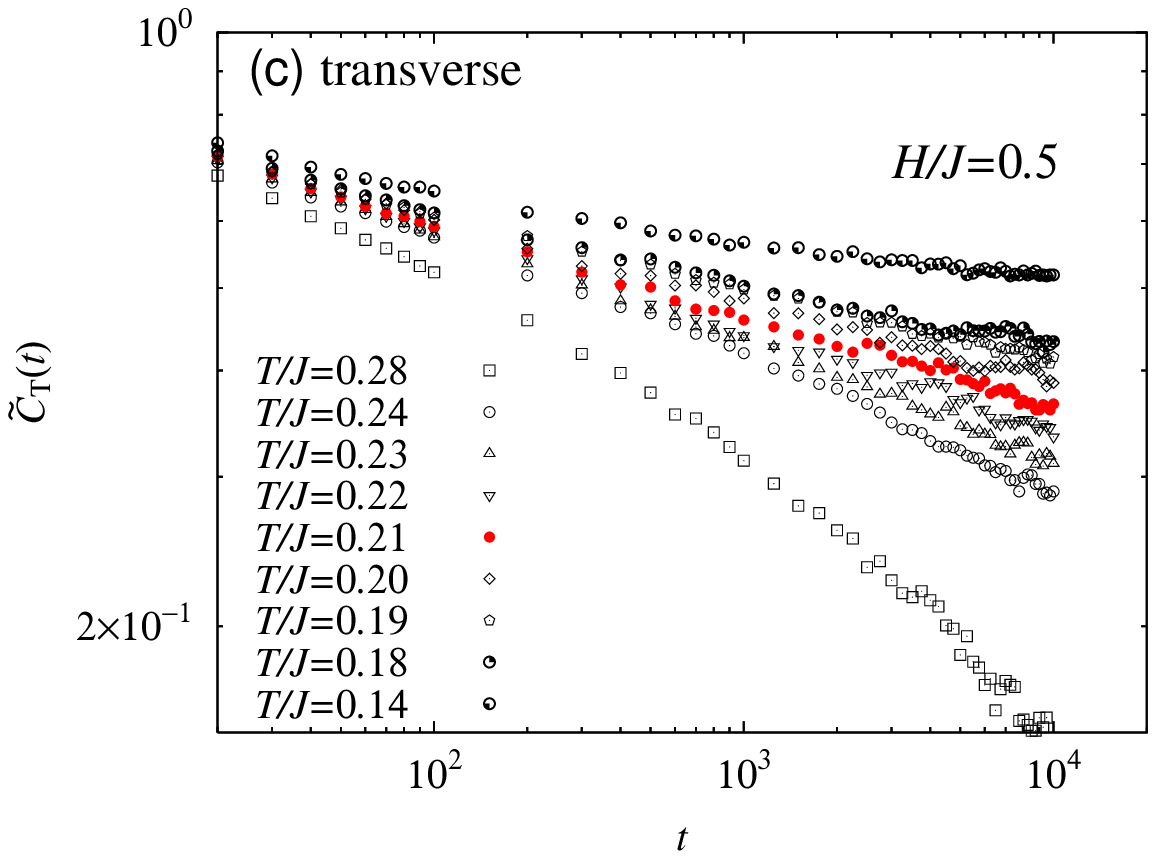} &
      \includegraphics[scale=0.7]{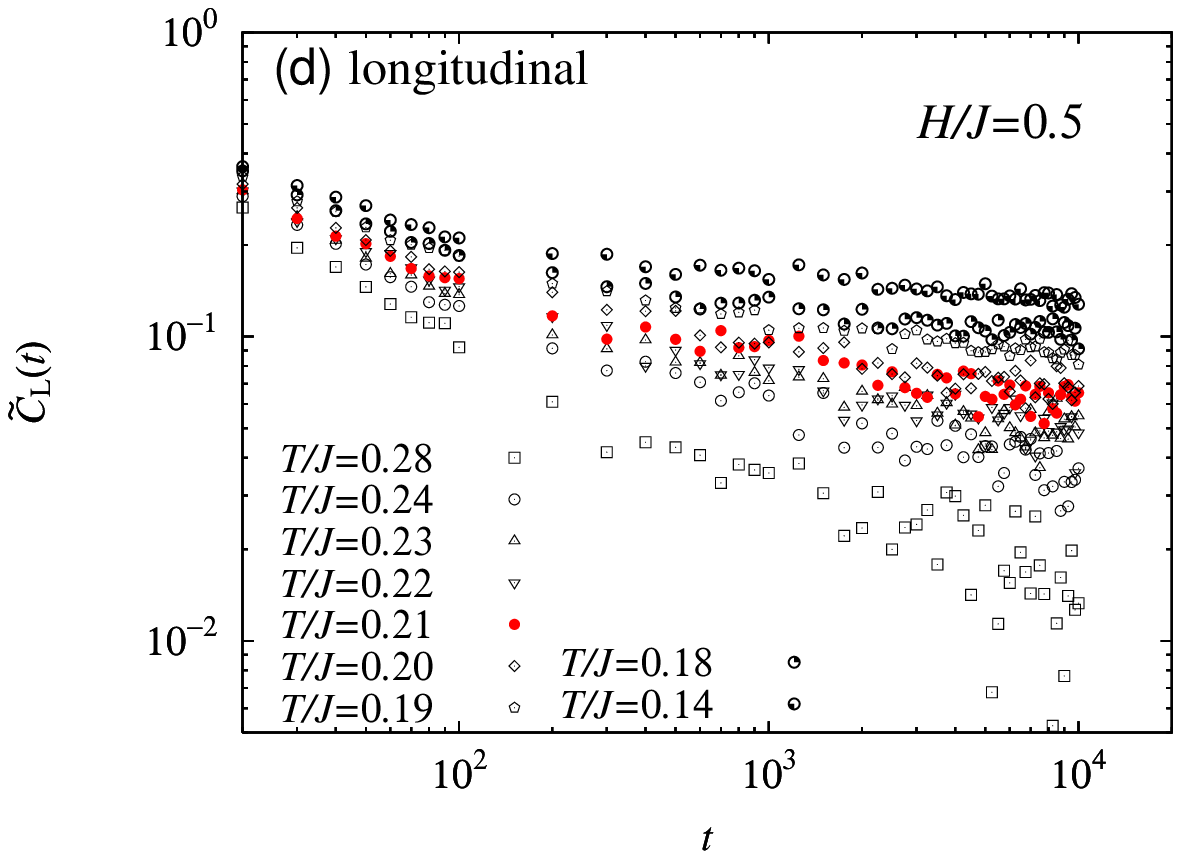}
    \end{tabular}
\caption{(Color online) 
Temporal decay of the spin autocorrelation
functions: (a) the transverse one $\tilde{C}_\mathrm{T}(t)$ for $H/J=0.05$,
(b) the longitudinal one $\tilde{C}_\mathrm{L}(t)$ for $H/J=0.05$,
(c) the transverse one for $H/J=0.5$, and
(d) the longitudinal one for $H/J=0.5$.
The lattice size is $L=16$. The data at $T=T_g$
are given in red (by filled symbols).
In the figures,
in order to check the finite size effect,
the data of $L=20$ are plotted with lines
at temperatures $T/J=0.18$-$0.24$ with an interval of 0.01.
The estimated transition
temperatures is $T_g/J\simeq 0.21$ both for $H/J=0.05$
and $H/J=0.5$.
}
    \label{fig_time_CT_D005}
  \end{center}
\end{figure}

In Figs.\ref{fig_time_CT_D005},
we show for the fields $H/J=0.05$ and 0.5 
the MC time dependence of the  autocorrelation
functions of the transverse component of the spin $\tilde{C}_\mathrm{T}(t)$, 
Figs.(a) and (c), and those
of the longitudinal component of the spin $\tilde{C}_\mathrm{L}$, 
Figs.(b) and (d). 
Finite-size effect evaluated from the difference between the $L=16$ and
$L=20$ data turns out to be rather small (but not completely negligible)
in the longitudinal component, whereas it is more
appreciable in the transverse component.
Although a finite-size effect is not
completely negligible here, the autocorrelation functions of the spin for
$H/J=0.05$ 
turn out to behave quite similarly to those of the chirality.
Namely, both
$\tilde{C}_\mathrm{L}(t)$ and $\tilde{C}_\mathrm{T}(t)$ exhibit either a
down-bending or an up-bending behavior depending on whether the temperature is
higher or lower than the borderline value $T/J\simeq 0.21$,
while just at this borderline temperature
a straight-line behavior corresponding a power-law decay is observed.
This indicates that the spin exhibits an RSB
transition at the same temperature where the chirality exhibits an RSB
transition.
The simultaneous occurrence of the spin and the chirality orderings
is quite natural
in the presence of the random anisotropy. 

For the field $H/J=0.5$, the transverse 
component $\tilde{C}_\mathrm{T}(t)$ shown in Fig.3(c) exhibits a rather
clear up-bending/down-bending behavior
with the borderline temperature $T/J\simeq 0.20-0.21$, which might be
compared with the less
clear behavior of $\tilde{C}_\chi(t)$ where the transition temperature appears
to lie somewhere between $T/J\simeq 0.21-0.24$. 
Since the transition here is
expected to be a simultaneous spin and chiral transition, we now estimate
the transition temperature of the field $H/J=0.5$ to be $T_g=0.21\pm 0.02$.
Meanwhile, the longitudinal autocorrelation function $\tilde{C}_\mathrm{L}(t)$
for $H/J=0.5$ exhibits a much noisier behavior as shown in Fig.3(d). This is 
because, in large fields, the second term of Eq.(30), 
$[\thermav{ q_{{\rm L}}}]$, 
becomes large, while the longitudinal spin correlation function itself becomes
small in magnitude which is obtained as a difference between the two large
numbers. Note that the up-bending behavior is discernible at 
a short time $t\simeq 10^2$, which is also observed in the chiral 
autocorrelation in a less pronounced manner. In any case, 
due to the noisiness, it seems not possible to identify the transition point 
from Fig.3(d).

From the behaviors of the chiral and the spin autocorrelation functions
shown above,
we conclude that
the spin and the chirality exhibit an RSB transition  simultaneously at
a finite temperature $T_g/J=0.21\pm 0.02$  for both the fields $H/J=0.05$ 
and 0.5.
The estimated transition temperature $T_g/J=0.21(2)$
is lower than the transition temperature of the corresponding zero-field model
with the same magnitude of anisotropy, {\it i.e.\/}, 
$T_g/J\simeq 0.24$,\cite{MatsuHuku,HK02} 
indicating that applied magnetic fields suppress the spin-glass
(chiral-glass) ordering for weaker fields. Indeed, this is exactly the feature 
expected for the experimental AT-line.
Interestingly, the estimated transition temperature under fields, 
$T_g/J\simeq 0.21$,  comes very close
to the chiral-glass transition temperature
of the fully isotropic model in zero field, which was estimated to be 
$T_\mathrm{CG}/J\simeq 0.20$.\cite{HK02}
This is exactly the feature expected for the experimental GT-line.
Hence, the suppression of $T_g(H)$
due to weaker fields (AT-line) as well as the robustness of it
with respect to stronger fields (GT-line) are
consistent with the experimental observation for the weakly anisotropic
Heisenberg-like SGs.\cite{SGrev,Campbell99,Campbell02}

\subsection{Comparison with the autocorrelation functions of the 3D Ising SG 
in fields}

For comparison, we also calculate the autocorrelation
function of the spin  $\tilde{C}_\mathrm{L}(t)$
defined by \eqtag(\ref{def_time_CL})
for the 3D Ising SG with the $\pm J$ coupling for the
field $H/J=0.05$. Note that the Ising SG in fields shares the same
symmetry  property as the anisotropic Heisenberg SG in fields, {\it i.e.\/},
the absence of any global symmetry. Nevertheless, {\it the Ising SG 
does not possess 
any chiral degree of freedom\/}. Thus, the question of whether the 3D Ising SG
behaves either similarly or differently from the weakly anisotropic 
Heisenberg SG would be of special interest.

\begin{figure}[ht]
  \leavevmode
  \begin{center}
    \begin{tabular}{ll}
      \includegraphics[scale=0.7]{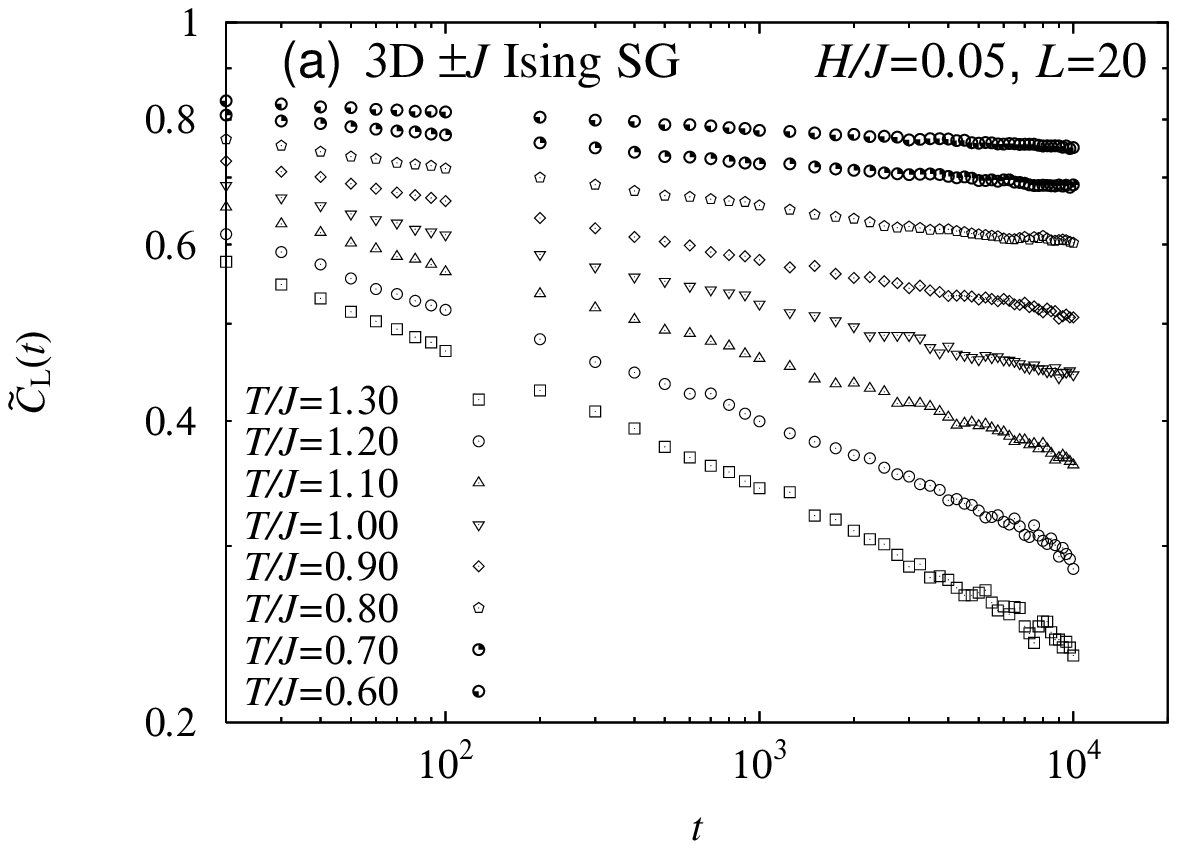}
      \includegraphics[scale=0.7]{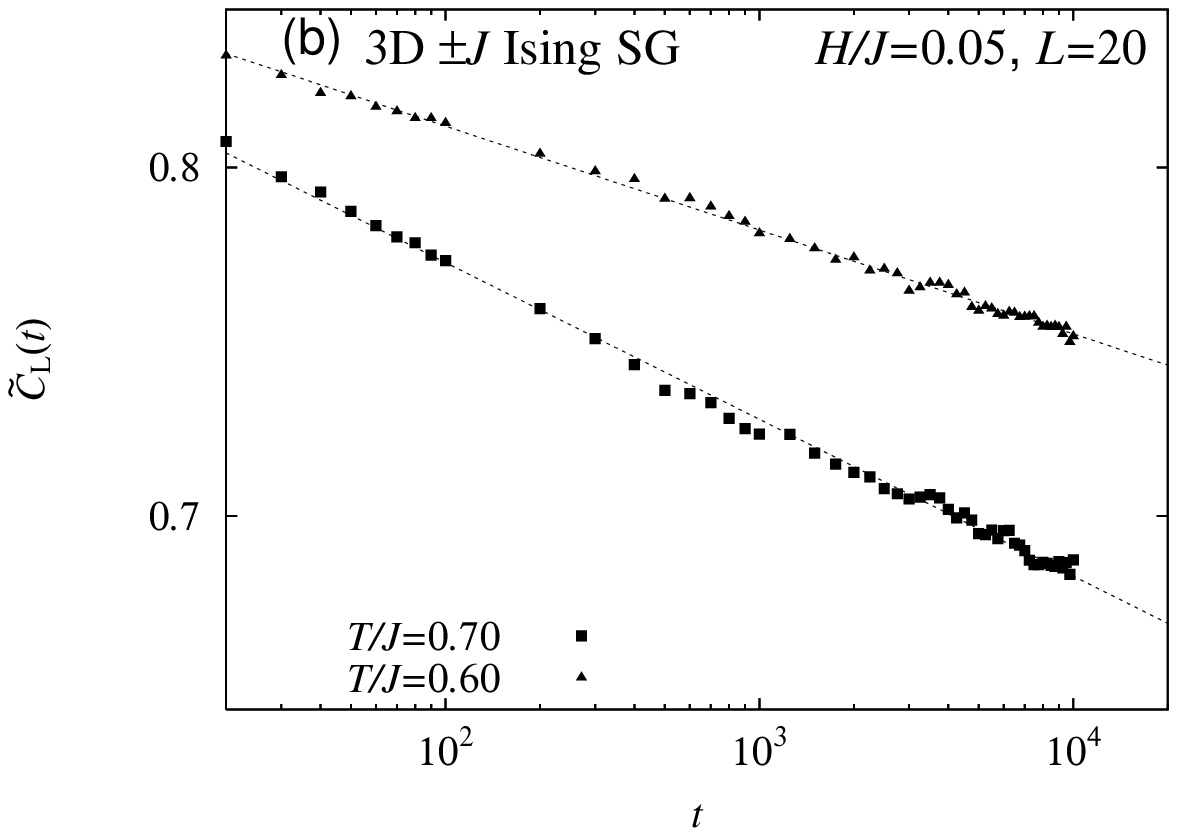}
    \end{tabular}
    \caption{
    Temporal decay of the spin autocorrelation
    function, defined by \eqtag(\ref{def_time_CL}),
    of the $\pm J$ 3D Ising SG in a magnetic field of $H/J=0.05$.
    The system size is $L=20$ averaged over 100 samples.
    In (a), the data  are plotted in the temperature range
    $T/J=0.60-1.3$ with an interval of 0.1.
    The data of the two lowest temperatures $T/J=0.60$ and 0.70
    are shown in (b),
    together with the fitted straight lines. Even at the lowest temperature
    studied, no up-bending behavior is observed.}
    \label{fig_time_C_3DISG_H005}
  \end{center}
\end{figure}

The data of $\tilde{C}_\mathrm{L}(t)$ of the Ising SG are
shown in Fig.\ref{fig_time_C_3DISG_H005} for the size $L=20$.
As can be seen from figure,
the behavior of $\tilde{C}_\mathrm{L}(t)$ of the Ising SG
differs significantly from that of the weakly anisotropic
Heisenberg SG.
Although the temperature range studied is as low
as about 60\%
of the zero-field transition temperature $T_g(H=0)/J\simeq 1.1$, which
is expected to be deep in the ordered state
according to a tentative estimate of Ref.\cite{Cruz03},
no clear up-bending behavior as observed in  the weakly anisotropic
Heisenberg SG is observed here. Instead,
$\tilde{q}^{(2)}_\mathrm{L}(t)$ and
$\tilde{C}_\mathrm{L}(t)$  persistently exhibit
an almost linear behavior even at the lowest temperature studied 
$T/J=0.6$.
This is illustrated in Fig.4(b) where the $\tilde{C}_\mathrm{L}(t)$ data
at the two lowest temperatures studied are shown. 
A comparison of the $L=20$ data with the $L=16$ data
indicates that some amount of finite-size effect still remains.
Nevertheless, an almost linear behavior without any discernible up-bending
tendency is robustly observed in common both for $L=16$ and $L=20$, suggesting
that this feature is a bulk property.

\subsection{The Binder ratio of the chirality}

\begin{figure}[ht]
  \leavevmode
  \begin{center}
    \includegraphics{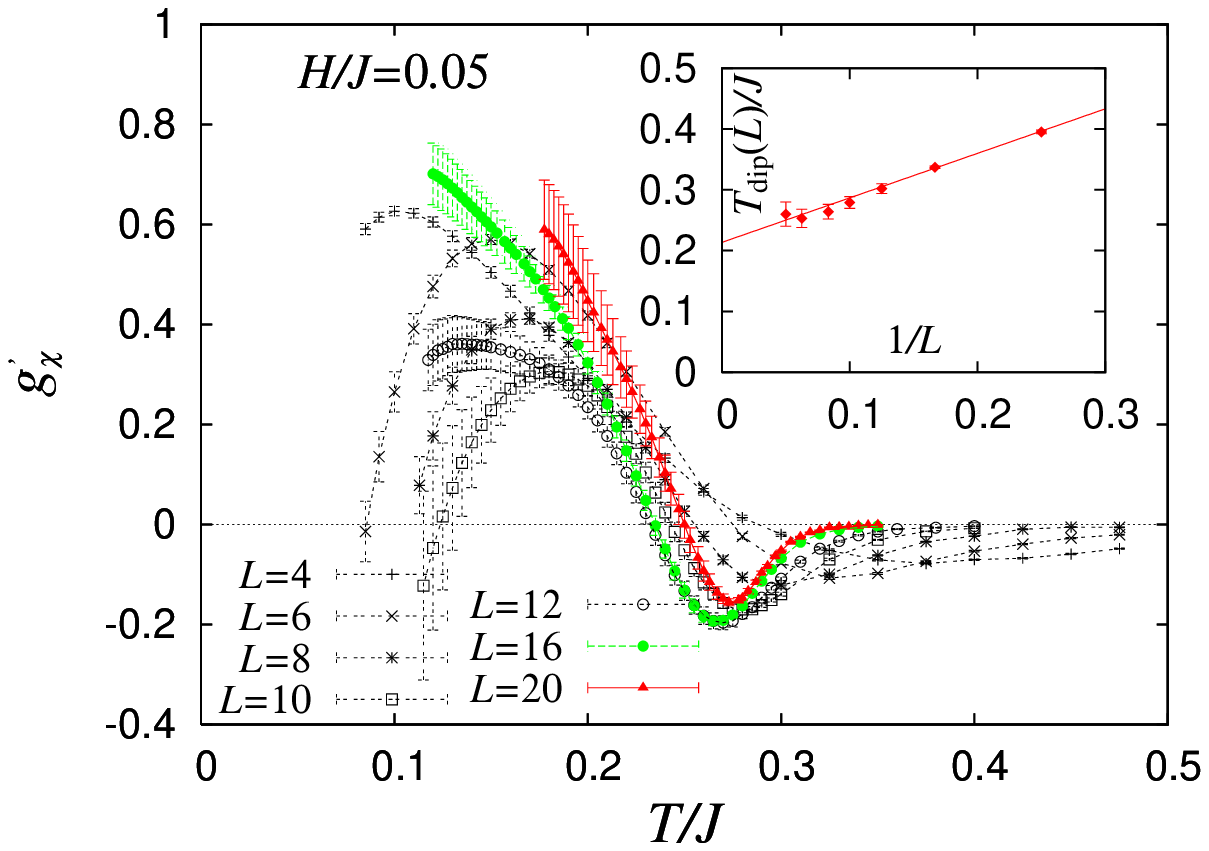}
\caption{(Color online) 
The temperature and size dependence of the
Binder ratio of the chirality $g^\prime_{\chi}$ for a field
$H/J=0.05$.
In the inset, we show
the reciprocal size dependence of the dip temperature of
$g^\prime_{\chi}$.
Dashed line represents a linear fit of the data.
The bulk chiral-glass transition temperature is estimated to be
$T_g/J=0.22(2)$.}
\label{fig_gx2_D005}
\end{center}
\end{figure}

In Fig.\ref{fig_gx2_D005}, we show the temperature- and size- dependence
of the chiral Binder ratio $g^{\prime}_{\chi}$ for the field $H/J=0.05$.
The data of $g^{\prime}_{\chi}$ has a negative dip at a size-dependent
temperature $T=T_{\rm dip}(L)$, which, with increasing $L$, tends to  deepen
and shift to lower temperatures.
The existence of a
persistent negative dip with increasing
depth, is a sign of a phase transition occurring at
$T=T_{{\rm dip}}(L=\infty)$. By extrapolating $T_{{\rm dip}}(L)$ to $L=\infty$,
as shown in the inset of Fig.\ref{fig_gx2_D005}, the bulk chiral-glass 
transition temperature is estimated
to be $T_g/J=0.22(2)$.
The estimated transition temperature agrees well with the
corresponding estimate based on the autocorrelation functions given above.

\begin{figure}[ht]
\begin{center}
\includegraphics[scale=1.5]{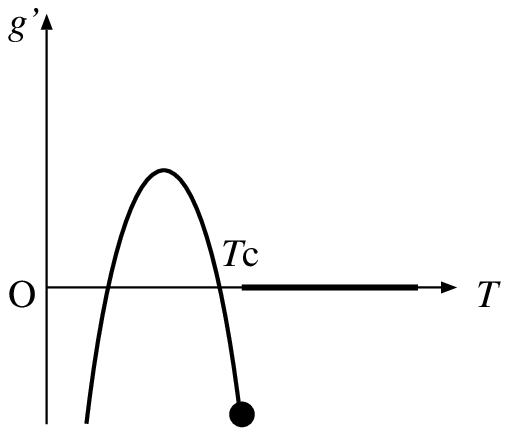}
\caption{
A sketch of the typical temperature dependence of the
Binder ratio $g^\prime$ in the thermodynamic limit, which is
expected in a system without a reflection symmetry exhibiting 
a  one-step RSB transition.}
\label{fig_gx_1stRSB}
\end{center}
\end{figure}

As argued in Ref.\cite{HK00R,ImaKawa02,HK00} for the isotropic case, 
the existence of a
persistent negative dip is a sign of a one-step-like RSB transition.
In the present anisotropic model under fields, there no longer exists a
global spin-reflection symmetry in contrast to the isotropic model
under fields.
Hence, the behavior expected for $g^{\prime}_{\chi}$ here
might be the one for the one-step RSB system {\it 
without a reflection symmetry\/}.
Such a system was theoretically analyzed in Ref.\cite{OPF}, 
where the behavior of
the Binder ratio $g_\chi'$ in the thermodynamic limit
was reported as shown in Fig.\ref{fig_gx_1stRSB}. 
\cite{OPF} Indeed, the overall behavior of our present
$g^{\prime}_{\chi}$ shown in Fig.\ref{fig_gx_1stRSB} 
seems consistent with such a behavior.




\subsection{The Binder ratio of the spin}

\begin{figure}[ht]
  \leavevmode
  \begin{center}
    \begin{tabular}{ll}
      \includegraphics[scale=0.7]{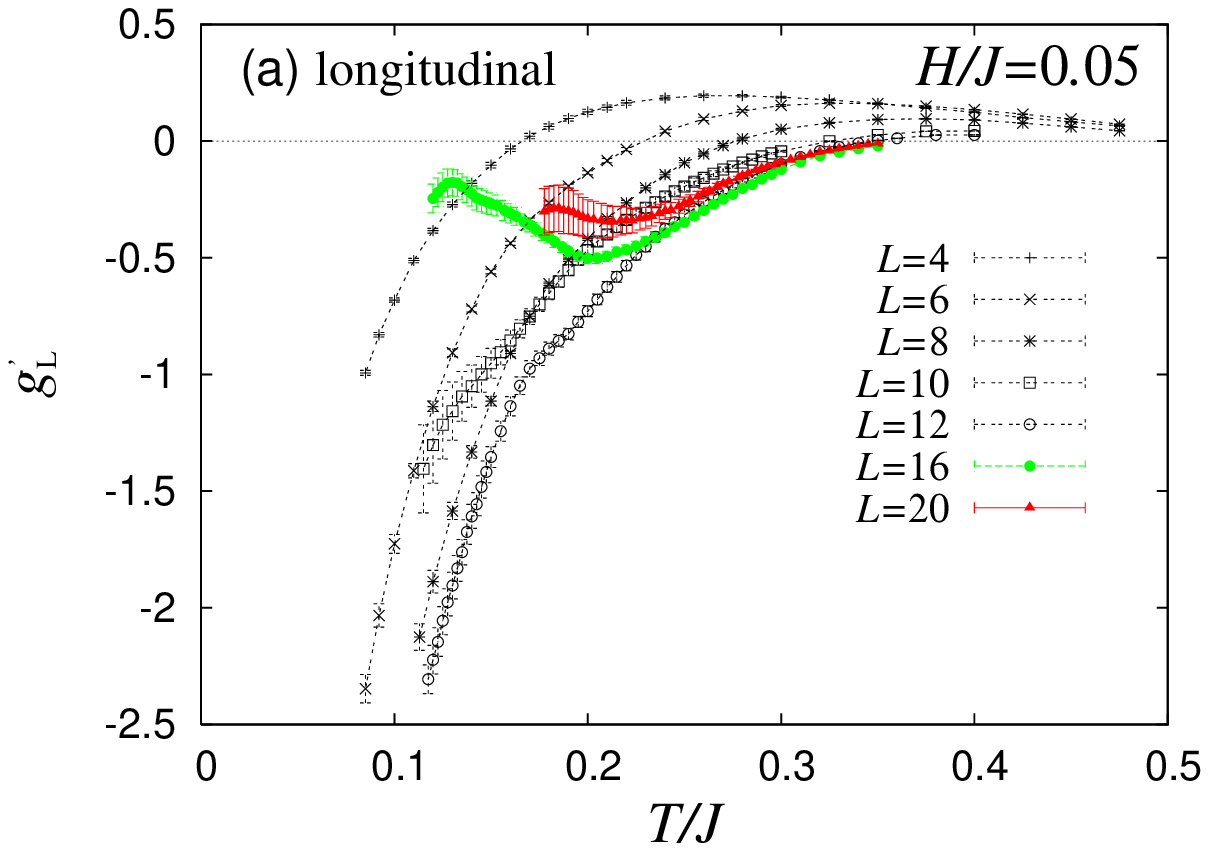} &
      \includegraphics[scale=0.7]{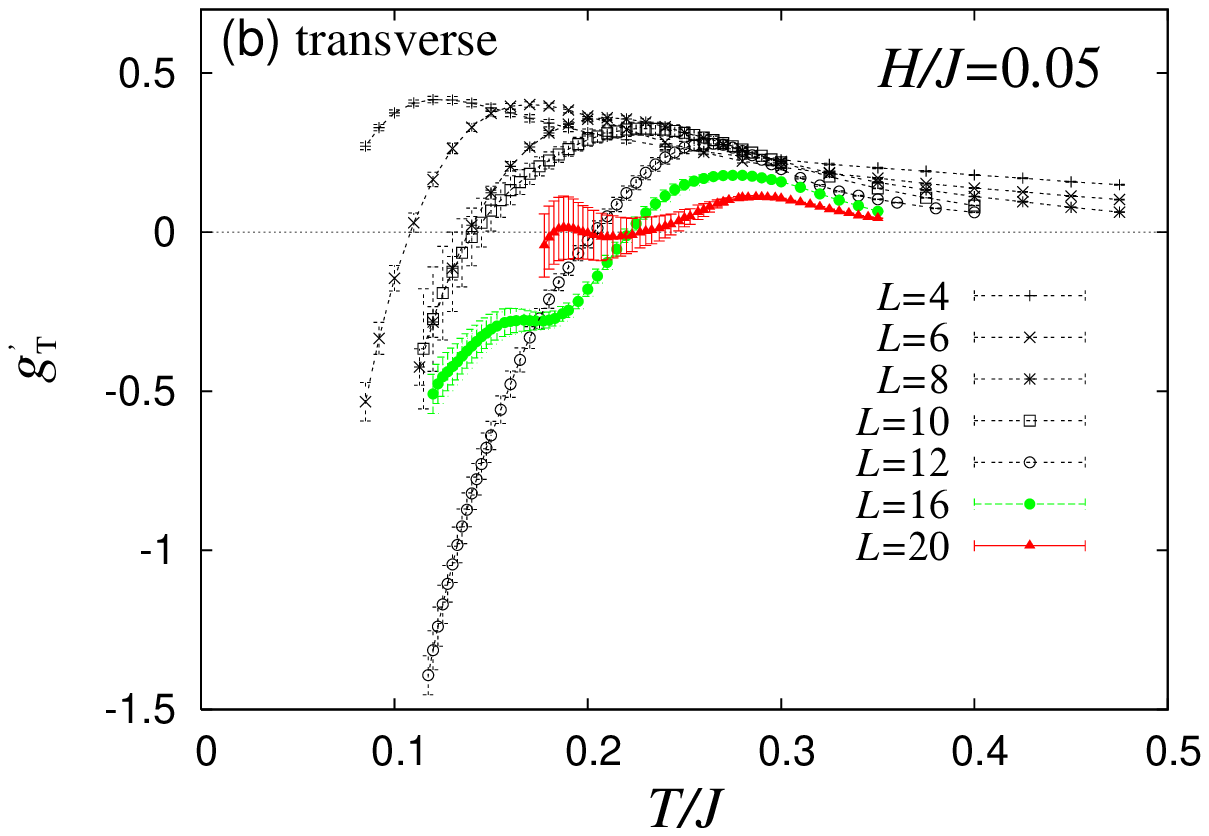}
    \end{tabular}
\caption{(Color online) 
The temperature and size dependence of the
Binder ratio of (a) the longitudinal component of the spin $g^{\prime}_\mathrm{L}$, and of
(b) the transverse component of the spin $g^{\prime}_\mathrm{T}$, 
for a field $H/J=0.05$.}
\label{fig_gL2_D005}
\end{center}
\end{figure}

In Fig.\ref{fig_gL2_D005}(a),
we show the temperature and size dependence of the Binder ratio
of the longitudinal component of the spin $g_\mathrm{L}^{\prime}$
for the field $H/J=0.05$.
Although $g_\mathrm{L}^{\prime}$ for smaller sizes $L\leq 12$
does not show any characteristic feature suggestive of a phase transition,
the one for larger sizes $L\geq 16$ tends to exhibit a negative dip  similar to the one as observed in the chiral Binder ratio $g_\chi^{\prime}$.
Then, for large enough $L$, the overall shape of $g_\mathrm{L}^{\prime}$ 
would be 
similar to that of $g_\chi^{\prime}$, although the one for smaller $L$ is 
very different.

In Fig.7(b),
we show the temperature and size dependence of the Binder ratio
of the transverse component of the spin $g_\mathrm{T}^{\prime}$
for the field $H/J=0.05$.
Although $g^\prime_T$ for smaller sizes $L\leq 12$ exhibits a
single maximum with a crossing point occurring around $T/J\simeq 0.28$,
the one  for larger sizes $L\geq 16$  exhibits double maxima.
The apparent crossing point observed for smaller sizes at $T/J\simeq 0.28$ 
disappears for larger lattices,
indicating
that it does not correspond to a true phase transition point.
Among the two maxima of $g^\prime_\mathrm{T}$ observed for $L\geq 16$,
the one at a higher temperature is gradually suppressed with increasing $L$,
while the one at a lower temperature tends to be enhanced.
Presumably, in the $L\rightarrow \infty$ limit, the peak at a higher
temperature will disappear,
while the peak at a lower temperature
will survive. Then, for large enough $L$, the overall shape of 
$g_\mathrm{T}^{\prime}$ would be 
similar to that of $g_\chi^{\prime}$ and $g^\prime_\mathrm{L}$, 
although the one for smaller $L$ is 
very different. Our observation here that the Binder ratio of the chirality, 
$g^\prime_\chi$, and those of the spin, $g^\prime_\mathrm{L}$ and
$g^\prime_\mathrm{T}$, asymptotically show mutually similar behavior, which 
resembles the one
depicted in Fig.6, seems consistent with the spin-chirality 
decoupling-recoupling scenario. Indeed, the recoupling length-scale
estimated in Ref.\cite{HK02,KawaLi01}, $L_\times \simeq 20$, is consistent 
with the observed behavior.

\subsection{Overlap distribution function of the chirality}

\begin{figure}[ht]
  \leavevmode
  \begin{center}
    \begin{tabular}{ll}
      \includegraphics[scale=0.7]{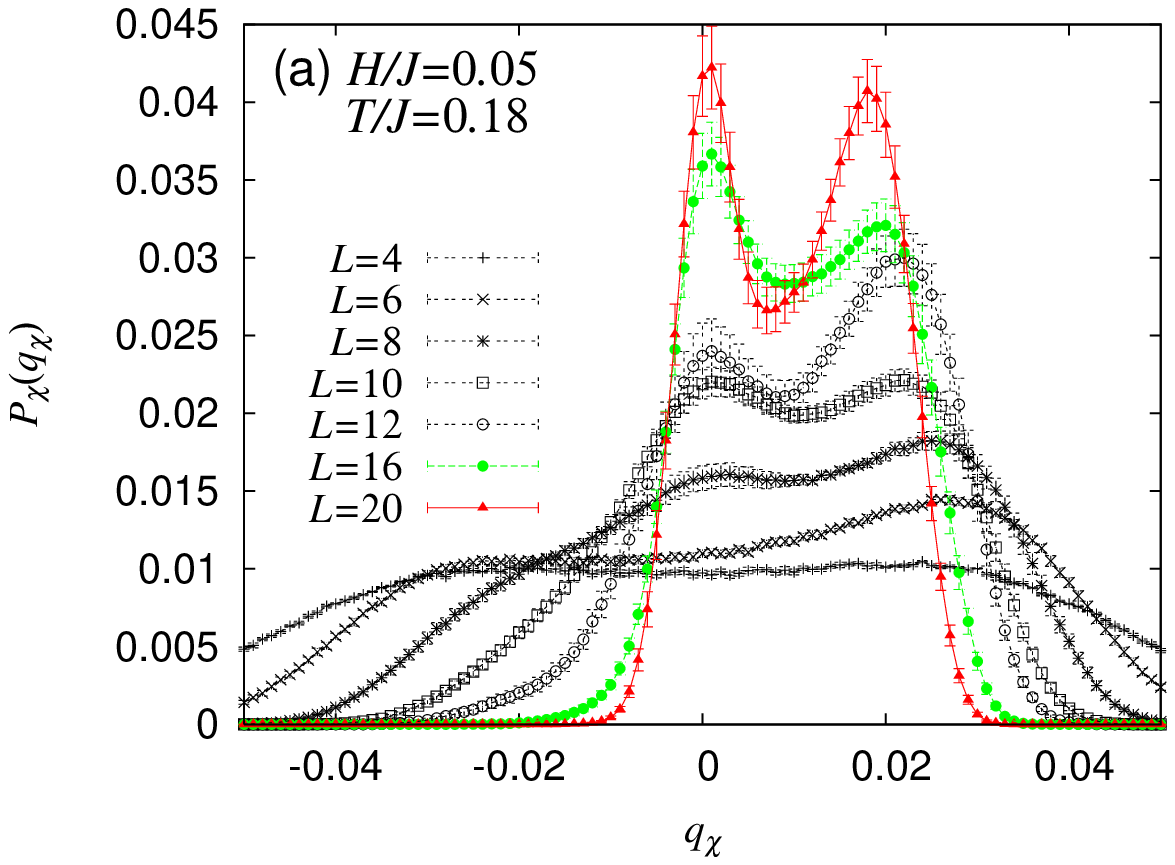} &
      \includegraphics[scale=0.7]{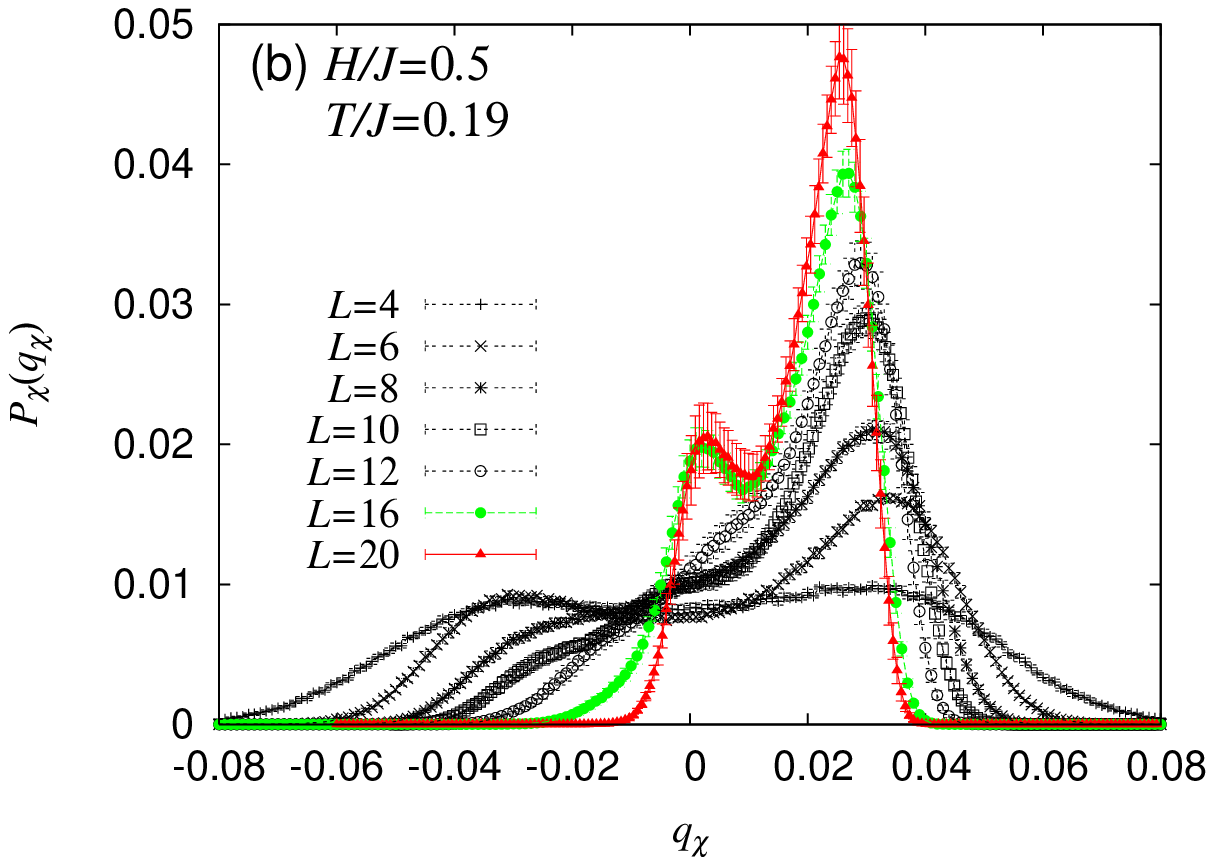}
    \end{tabular}
    \begin{tabular}{ll}
      \includegraphics[scale=0.7]{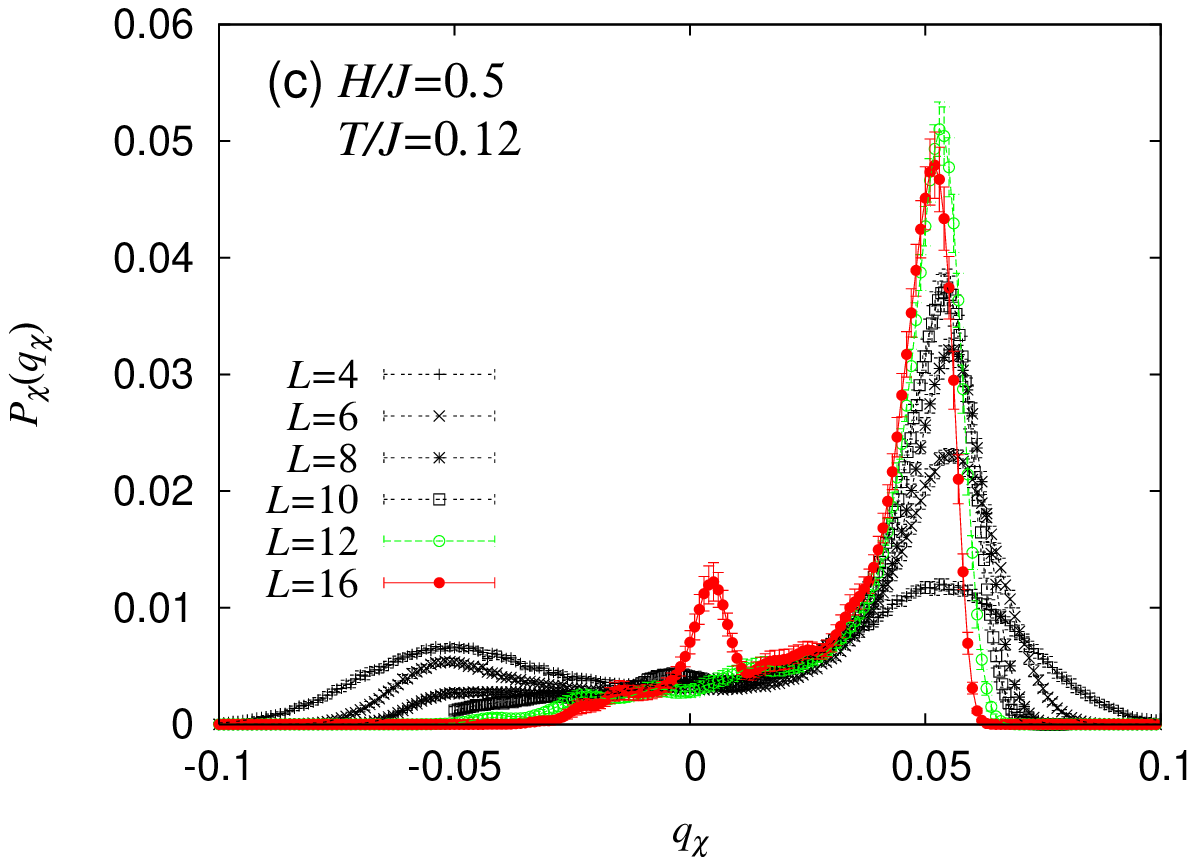} &
      \includegraphics[scale=0.7]{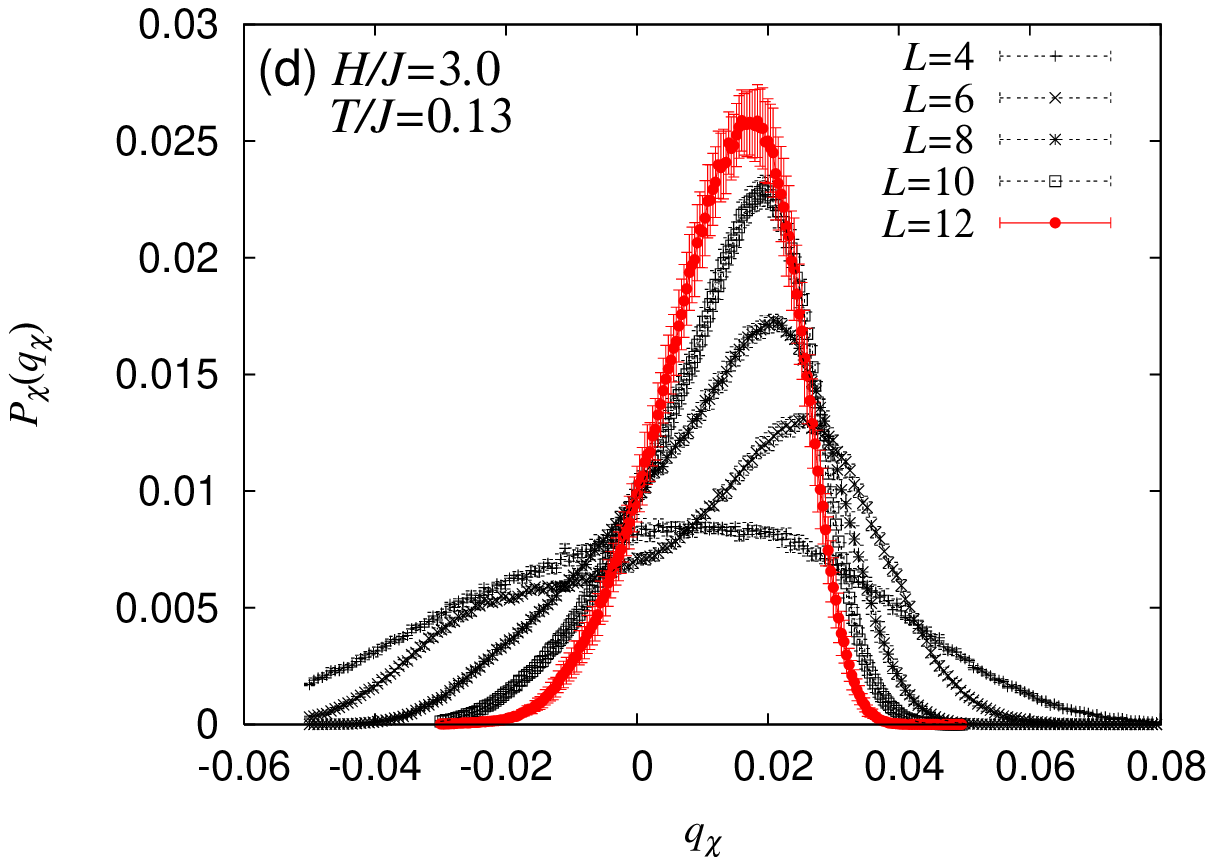}
    \end{tabular}
\caption{(Color online) 
The chiral-overlap distribution functions for the field and the
temperature,
(a) $H/J=0.05$ and $T/J=0.18$,
(b) $H/J=0.5$ and $T/J=0.19$, (c) $H/J=0.5$ and $T/J=0.12$, and
(d) $H/J=3.0$ and $T/J=0.13$.
The transition temperature is 
$T_g/J \simeq 0.21$ for both cases of $H/J=0.05$ and 0.5, while it is 
lower than $0.13$ for $H/J=3$.
}
    \label{fig_Pqx_D005}
  \end{center}
\end{figure}

We show in Figs.\ref{fig_Pqx_D005} the
size dependence of the overlap distribution function of the chirality, 
$P_{\chi}(q_{\chi})$, 
for several cases, {\it i.e.\/},
(a) $H/J=0.05$ and $T/J=0.18$, (b) $H/J=0.5$ and $T/J=0.19$, (c) $H/J=0.5$ 
and $T/J=0.12$, and (d) $H/J=3.0$ and $T/J=0.13$.

In the case $H/J=0.05$,
in addition to a primary peak
corresponding to $q_{\chi}=q_{\chi}^{\rm EA}>0$, 
which grows and sharpens with increasing $L$,
there appears a second peak at around $q_{\chi}=0$,
which also grows and sharpens with increasing $L$.
The existence of two distinct
peaks, both growing and sharpening with increasing $L$,
is a clear indication of the occurrence of RSB.
As reported in Ref.\cite{HK00R},
$P_{\chi}(q_{\chi})$ in zero field exhibits a feature of a
one-step-like RSB, {\it i.e.\/},
a central peak at $q_\chi=0$
coexisting with self-overlap peaks at $q_\chi=\pm q_\chi^{{\rm EA}}$.
The $P_{\chi}(q_{\chi})$ observed here may be regarded
as the in-field counterpart of the zero-field $P_{\chi}(q_{\chi})$
with a feature of such a one-step-like RSB.
Indeed, if one closely looks at $P_{\chi}(q_{\chi})$ 
shown in Fig.\ref{fig_Pqx_D005}(a), one sees that a broad peak,
which is a remnant of the $q_{\chi}=-q_{\chi}^{\rm EA}$ peak of the zero-field 
model,
is discernible for smaller sizes $L\leq 8$,
whereas, for larger sizes, this $q_{\chi}=-q_{\chi}^{\rm EA}$ peak disappears 
and the $q_{\chi}\simeq 0$ peak begins to grow. Interestingly, for the size
$L=8$, $P_{\chi}(q_{\chi})$ possesses {\it three\/} broad peaks, at around 
$q_{\chi}=\pm q_{\chi}^{\rm EA}$ and $q_\chi =0$. Such a three-peak structure
is rarely seen in a system exhibiting the full RSB, and gives a further
indication that the RSB occurring here is the in-field counterpart
of the one-step-like RSB.

For $H/J=0.5$, $P_{\chi}(q_{\chi})$ shows a 
similar behavior as that for $H/J=0.05$. Namely, for larger $L$,
it exhibits two distinct
peaks, both growing and sharpening with increasing $L$, 
whereas it exhibits three broad peaks for an intermediate $L$. As compared
with the $H/J=0.05$ case, 
the second peak is located slightly off $q_{\chi}=0$, reflecting  
the fact that the higher field breaks the 
$q_{\chi}
\leftrightarrow -q_{\chi}$ symmetry more strongly.
Anyway, our data shown in Figs.\ref{fig_Pqx_D005}(a)-(c) 
give a strong numerical support
that there indeed occurs a chiral-glass transition at a finite temperature
and that the chiral-glass ordered state accompanies a one-step-like RSB.

For the still higher field $H/J=3.0$,
in contrast to the cases of $H/J=0.05$ and 0.5,
the double-peak behavior of $P_\chi(q_\chi)$ is not observed even at 
the lowest temperature studied $T/J=0.13$.
Thus, for $H/J=3.0$,
no sign of RSB transition is observed down to this low temperature.

\subsection{Overlap distribution function of the spin}

\begin{figure}[ht]
  \leavevmode
  \begin{center}
    \begin{tabular}{ll}
      \includegraphics[scale=0.7]{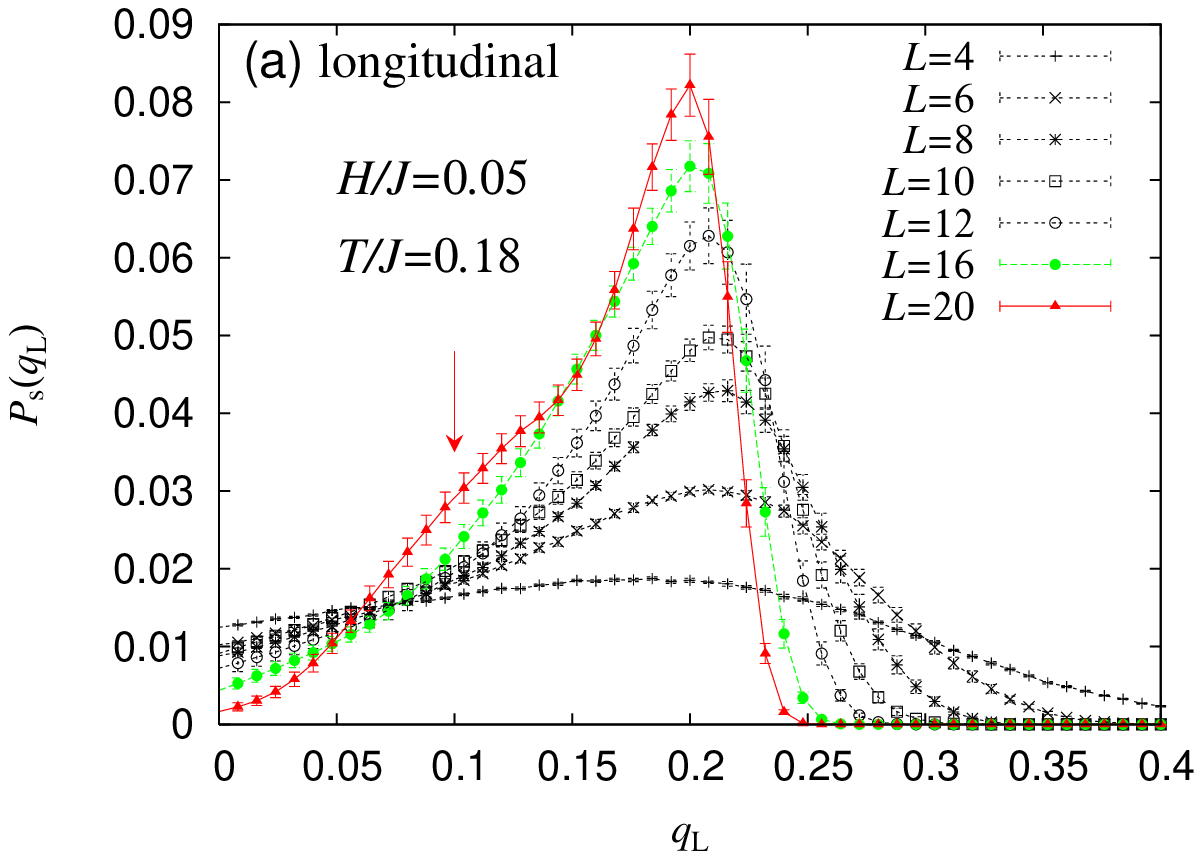} &
      \includegraphics[scale=0.7]{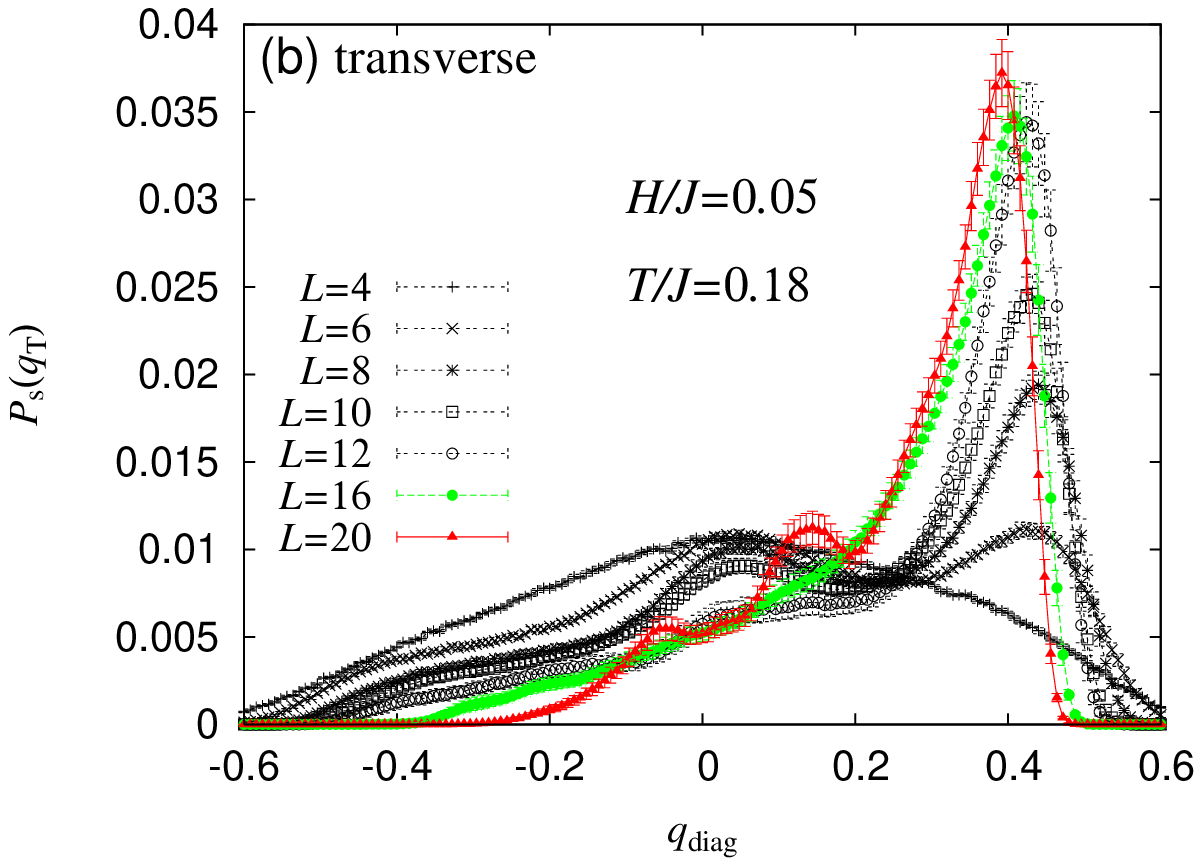}
    \end{tabular}
\caption{(Color online) 
The overlap distribution functions of (a) the longitudinal
component of the spin,
and of (b) the transverse component of the spin, for a field $H/J=0.05$ 
at a temperature $T/J=0.18$.
The transition temperature is
$T_g/J \simeq 0.21$. In Fig.(a), a sign
of the second peak just begins to emerge  for the largest
size $L=20$ at the arrow position.
}
    \label{fig_Pqz_D005}
  \end{center}
\end{figure}

We show in Figs.\ref{fig_Pqz_D005} the
size dependence of the overlap distribution function of (a) the longitudinal
component of the spin $P_\mathrm{s}(q_\mathrm{L})$, 
and of (b) the transverse
component of the spin $P_\mathrm{s}(q_\mathrm{T})$, 
for the field $H/J=0.05$ and at a temperature $T/J=0.18$. The longitudinal
spin-overlap distribution function 
$P_\mathrm{s}(q_\mathrm{L})$ for smaller sizes $L\leq 16$ possesses only a 
single growing peak at around $q_\mathrm{L}\simeq 0.2$, in apparent contrast to
the double-peak structure observed in the chiral-overlap distribution.
Quite interestingly, however,
for the  largest size studied $L=20$, one sees that
the second peak just begins to emerge at around $q_\mathrm{L}\simeq
0.1$ (see the arrow
in Fig.9(a)).
This second peak is reminiscent to the one observed in the 
chiral-overlap distribution function of Fig.8, 
though it appears here only for the largest
size in a less pronounced manner: It appears to be an echo of the strongly
diverging $q_\chi\simeq 0$ peak observed in the chiral-overlap distribution.
Our observation that the one-step RSB-like structure of the
overlap distribution appears in the chiral sector from smaller sizes,
while it appears in the spin sector in a less pronounced manner 
only for larger sizes, suggests that the order parameter of the present
one-step-RSB transition
might be the chirality, rather than the spin. Again, this observation
is fully consistent with the spin-chirality  decoupling-recoupling 
scenario.\cite{Kawamura92,Kawamura98}

As shown in Fig.9(b),
the behavior of the transverse-spin-overlap distribution function
$P_\mathrm{s}(q_\mathrm{T})$ is more complex. Such a behavior
is certainly consistent with the occurrence of an RSB transition.

\subsection{Correlation lengths}

\begin{figure}[ht]
\leavevmode
  \begin{center}
    \begin{tabular}{l}
      \includegraphics[scale=0.7]{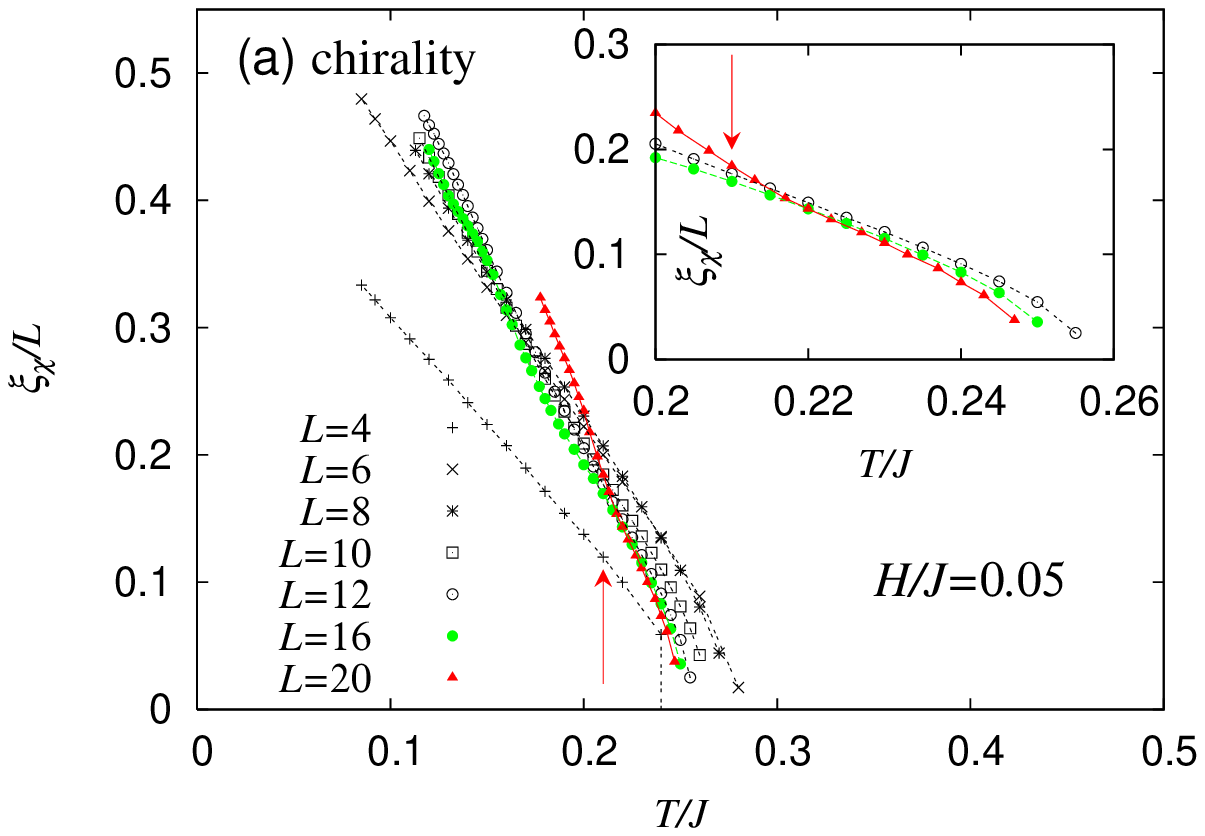}
    \end{tabular}
    \begin{tabular}{ll}
      \includegraphics[scale=0.7]{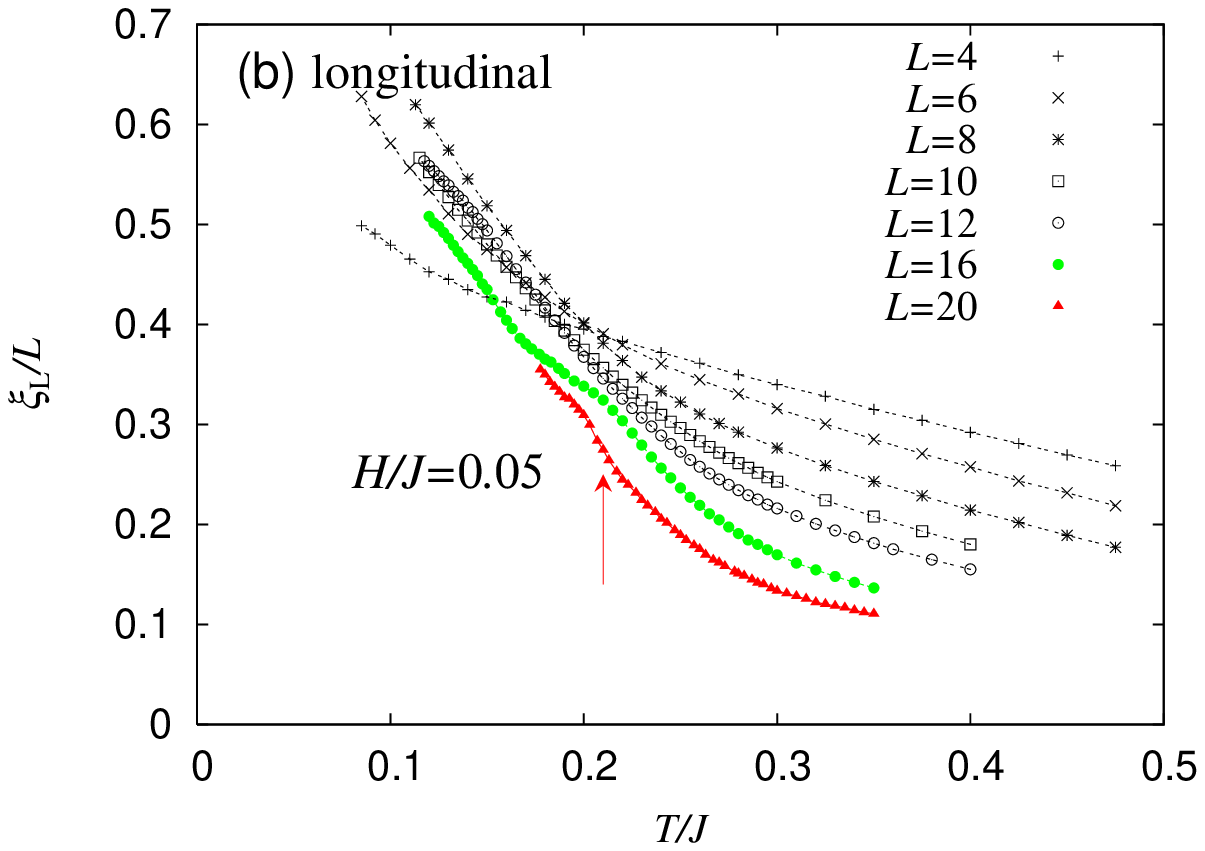} &
      \includegraphics[scale=0.7]{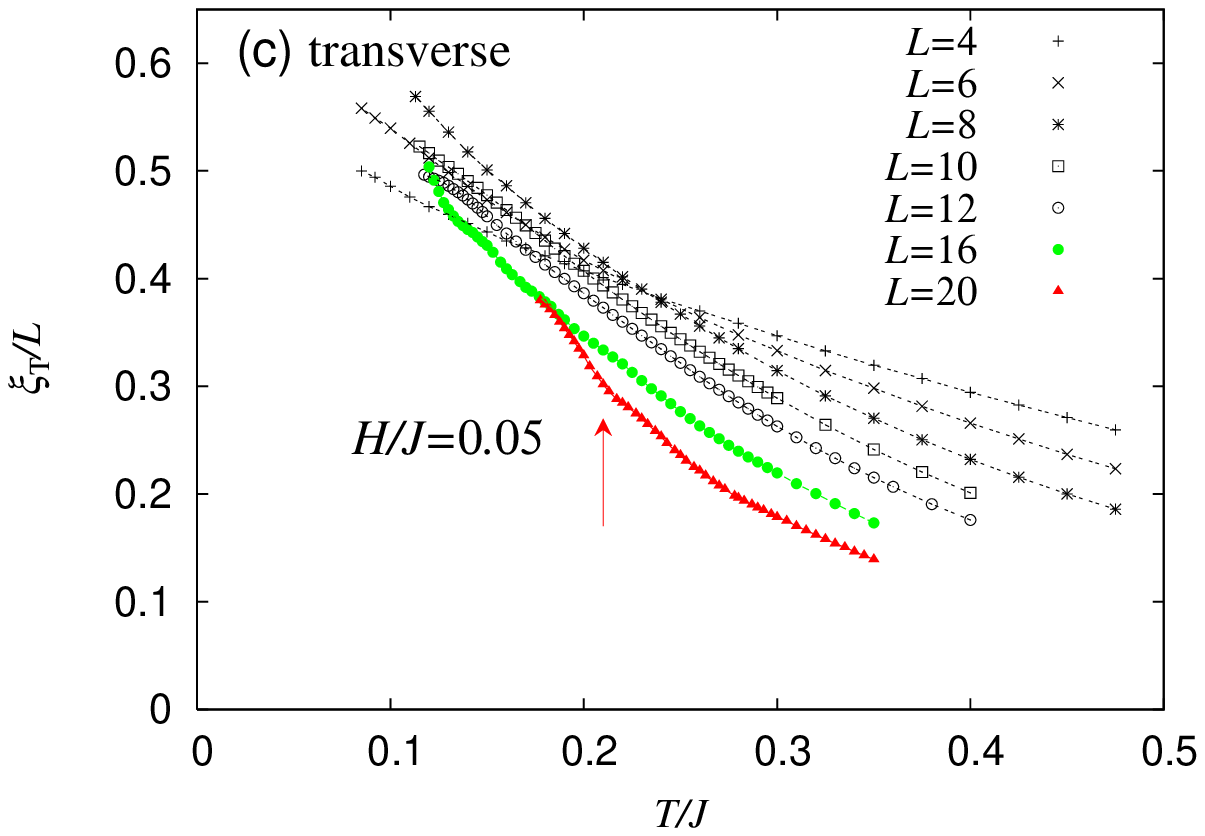}
    \end{tabular}
\caption{(Color online) 
The temperature and size dependence of the normalized
correlation lengths of
(a) the chirality, of (b) the longitudinal component of the spin,
and of (c) the transverse component of the spin,
for a field $H/J=0.05$. In Fig.(a),
a magnified figure is given in the inset
for the three largest sizes, $L=12$, $16$ and $20$.
The arrows in the figures indicate the locations of $T_g(H)$,
{\it i.e.\/}, $T_g/J=0.21$.}
    \label{fig_xi_x_D005}
 \end{center}
\end{figure}

We show in Figs.\ref{fig_xi_x_D005} the temperature and size dependence
of the normalized correlation lengths of (a) the chirality $\xi_\chi/L$, of
(b) the longitudinal component of the spin $\xi_{\rm L}/L$,
and of (c) the transverse component of the spin $\xi_{\rm T}/L$, 
for the field $H/J=0.05$.
The location  of $T_g(H)$ obtained from the autocorrelation functions are
displayed with arrows in the figures.

Concerning the chirality, as can be seen
from the inset of Fig.10(a).
a clear crossing behavior is observed at $T/J\simeq 0.22$ 
for larger sizes $L\geq 12$. 
The crossing temperature $T/J\simeq 0.21$ observed for larger sizes 
turns out to be close to our previous estimate of $T_g$.

Concerning  the longitudinal and the transverse components of the spin,
a crossing
is observed at $T_g/J\simeq 0.21$ for smaller sizes $L\leq 8$. For 
$L\geq 10$, however, $\xi/L$ decreases
with $L$ at any temperature studied, 
no longer exhibiting a crossing at $T_g/J\simeq 0.21$.
For even  larger sizes $L=16$ and $L=20$,  a
tendency of crossing re-appears at
around $T_g/J\simeq 0.21$, but now at a lower value of $\xi/L$.

Such a complex size dependence of the spin correlation lengths may
naturally be interpreted
by the spin-chirality decoupling-recoupling scenario in the following way: 
For smaller sizes, {\it i.e.\/}, at shorter length scales,
the spin and the chirality are trivially coupled in correlations
irrespective of the anisotropy,
so that the crossing in the sizes $L\leq 8$ may reflect this trivial
coupling at short length scales.
For larger sizes, {\it i.e.\/}, at long length scales, the spin is
``recoupled'' to the chirality via the anisotropy in a way different from the
trivial coupling at short length scales. Thus, the two different
types of crossing is expected in the normalized correlation lengths
for smaller and for larger sizes, 
which is exactly the behavior observed in Figs.10(b) and (c).
The characteristic length scale separating the coupling and the
recoupling regimes was estimated to be about $20$ lattice spacings.
\cite{HK02,KawaLi01}
Therefore, although we cannot simulate here the sizes larger than $L=20$
due to the lack of our computational capability, we do expect that
a clear crossing behavior will eventually set in for $L>20$.

\section{The critical properties and the phase diagram}

In this section, we further analyze the nature of the chiral-glass (spin-glass)
transition observed in the previous section
by means of a scaling analysis, and construct a magnetic
phase diagram of the model in the temperature - magnetic field plane.

We apply a dynamical scaling analysis
to the chiral autocorrelation functions
in order to estimate the critical
exponents of the transition. The
standard bulk dynamical scaling form is assumed for the 
autocorrelation functions, 
$\tilde{C}_\chi(t)$, $\tilde{C}_\mathrm{L}(t)$
and $\tilde{C}_\mathrm{T}(t)$,
\begin{equation}
\tilde{C}(t)\approx|(T-T_g)/J|^{\beta}f(t|(T-T_g)/J|^{z\nu})
\end{equation}
where $T_g$ is the transition temperature determined in the previous 
section, while $\beta$,
$\nu$ and $z$ refer to the order
parameter, the correlation-length and the dynamical exponents,
respectively.
The exponents $\beta$ and  $z\nu$ are to be
determined so that  a good data collapse is obtained in the
scaling plot. The quality of the scaling plot is judged by eyes.

\subsection{Dynamical scaling analysis of the chiral autocorrelation 
function}

\begin{figure}[ht]
\leavevmode
  \begin{center}
    \begin{tabular}{ll}
      \includegraphics[scale=1.2]{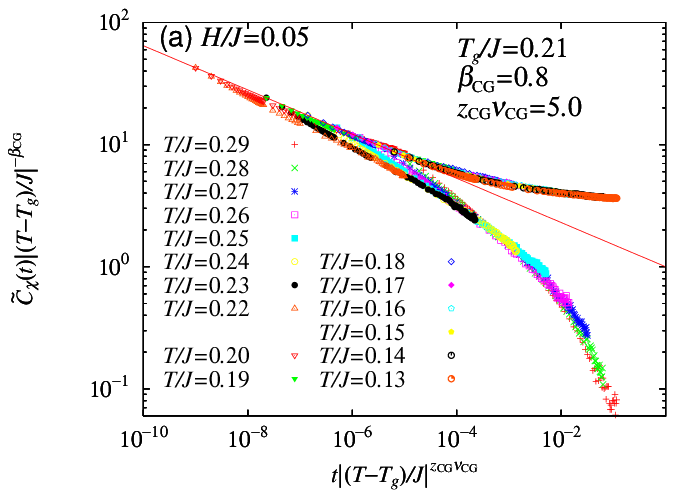} &
      \includegraphics[scale=1.2]{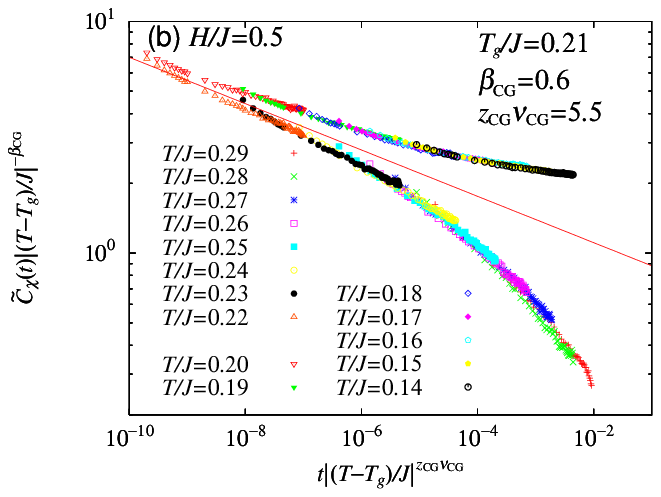}
    \end{tabular}
    \caption{(Color online) 
Dynamical scaling plots of the chiral autocorrelation function $\tilde{C}_\chi(t)$, 
    for fields (a) $H/J=0.05$ and (b) $H/J=0.5$.}
    \label{fig_DynSc_Cx_D005}
  \end{center}
\end{figure}

In Figs.\ref{fig_DynSc_Cx_D005},
we show the scaling plots of the chiral autocorrelation function 
$\tilde{C}_\chi(t)$
for the fields (a) $H/J=0.05$ and (b) $H/J=0.5$.
The transition temperature is fixed to be
$T_\mathrm{CG}/J=0.21$ both for $H/J=0.05$ and $H/J=0.5$, 
as  determined in the previous section.

As can be seen from the figures,
the chiral autocorrelation function
scales well both above and below $T_g$, with $\beta_\mathrm{CG}=0.8$
and $z_\mathrm{CG}\nu_\mathrm{CG}=5.0$ for $H/J=0.05$,
and with $\beta_\mathrm{CG}=0.7$
and $z_\mathrm{CG}\nu_\mathrm{CG}=5.5$ for $H/J=0.5$. Examining by eyes the
quality of the scaling plots with varying the assumed exponents values, we
finally quote $\beta_\mathrm{CG}=0.8\pm 0.2$
and $z_\mathrm{CG}\nu_\mathrm{CG}=5.0 \pm 1.0$ for $H/J=0.05$,
and $\beta_\mathrm{CG}=0.7 \pm 0.2$
and $z_\mathrm{CG}\nu_\mathrm{CG}=5.5 \pm 1.0$ for $H/J=0.5$.
The estimated chiral-glass exponents are
not far from the corresponding zero-field values 
$\beta_{{\rm CG}} \simeq 1.1$\cite{Kawamura98} and
$z_\mathrm{CG}\nu_\mathrm{CG}\simeq 4.5$. \cite{MatsuHuku} 

\subsection{Dynamical scaling analysis of the spin autocorrelation functions}

\begin{figure}[ht]
\leavevmode
  \begin{center}
    \begin{tabular}{ll}
      \includegraphics[scale=1.2]{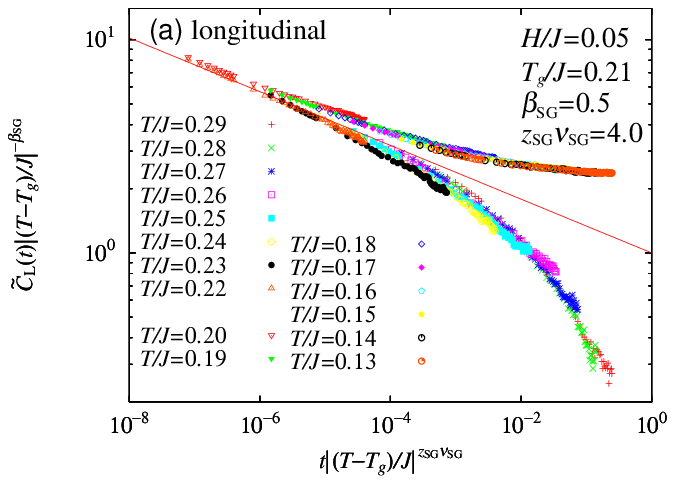} &
      \includegraphics[scale=1.2]{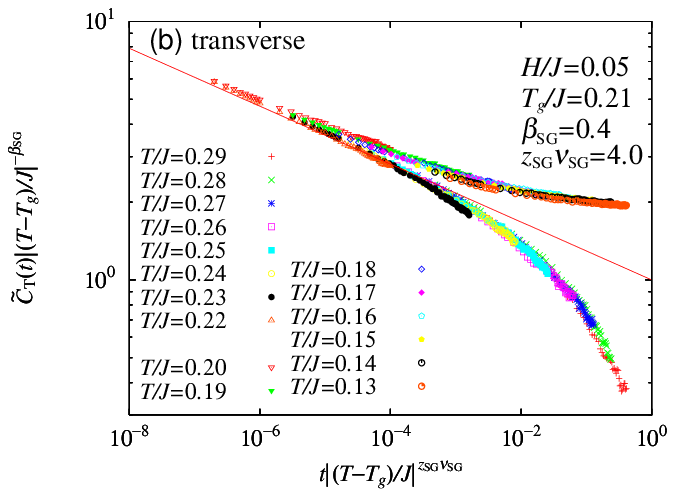}
    \end{tabular}
    \caption{(Color online)
    Dynamical scaling plots of the autocorrelation functions
    of (a) the longitudinal component of the spin $\tilde{C}_\mathrm{L}$,
    and of (b) the transverse component of the spin $\tilde{C}_\mathrm{L}$,
    for a field $H/J=0.05$.
    }
    \label{fig_DynSc_Cs_D005}
  \end{center}
\end{figure}

Next, we apply a dynamical scaling analysis
to the spin autocorrelation functions.
In Figs.\ref{fig_DynSc_Cs_D005},
we show the scaling plots both for (a) the longitudinal and 
for (b) the transverse components of the spin
for the field $H/J=0.05$.
The transition temperature is fixed to be
$T_g/J=0.21$.
In Ref.\cite{ImaKawa04}, 
we observed that the quality of the scaling plot of the spin
time-correlation function $q_\mathrm{L}^{(2)}$ was not as good as 
its chiral counterpart $q_\chi^{(2)}$. 
As argued in Ref.\cite{ImaKawa04}, 
if the chirality, rather than the spin, is the order parameter of the 
transition, 
in the time regime shorter than the recoupling time scale (estimated to be
$10^5\sim 10^6$ MCS), 
the spin time-correlation might not reach an asymptotic scaling regime
even when the chiral time-correlation reaches an asymptotic scaling regime.
Hence, although the scaling of $\tilde{C}_\mathrm{L}(t)$
and $\tilde{C}_\mathrm{T}(t)$ turns out to be fairly good as shown in Fig.12,
caution might be required here in
regarding the fitted values of $\beta_\mathrm{SG}$ and 
$z_\mathrm{SG}\nu_\mathrm{SG}$
as true asymptotic spin exponents, since 
our observation time window
is shorter than the expected recoupling time scale.

\subsection{Chiral-glass and spin-glass susceptibilities}

\begin{figure}[ht]
   \leavevmode
   \begin{center}
     \begin{tabular}{ll}
      \includegraphics[scale=0.7]{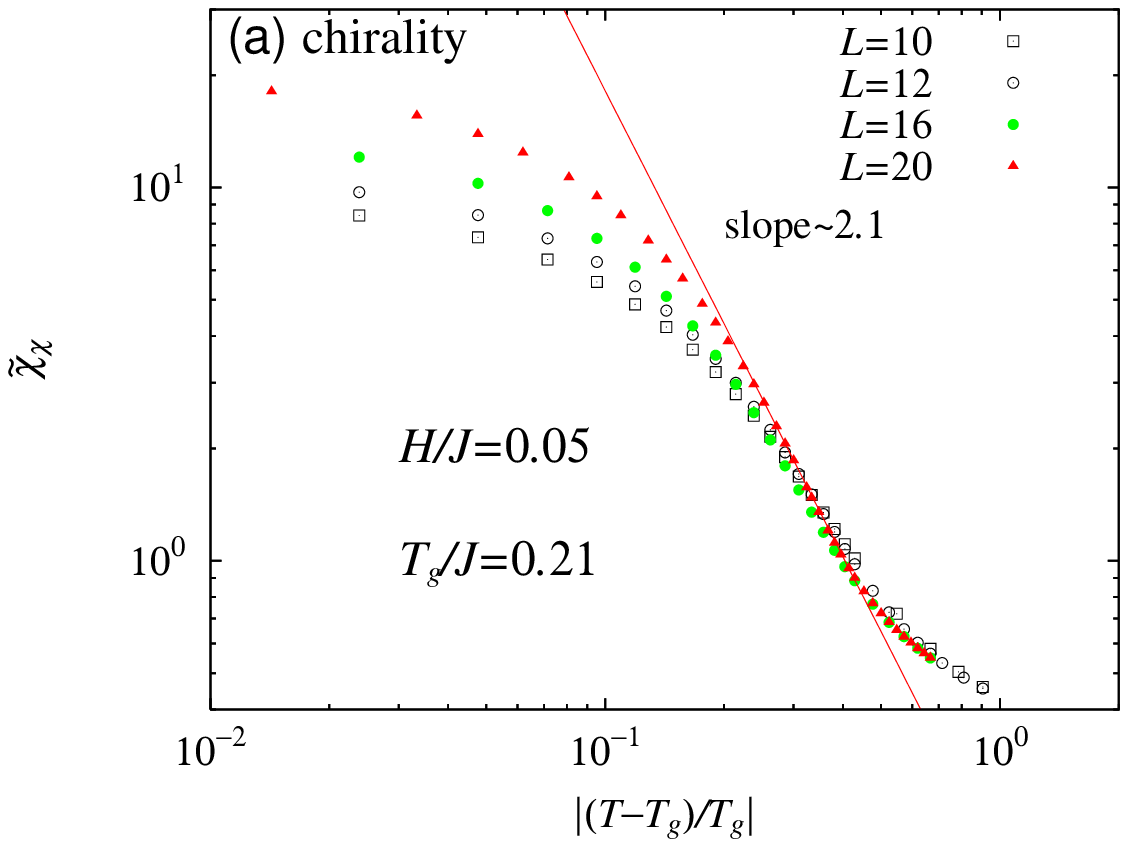} &
      \includegraphics[scale=0.7]{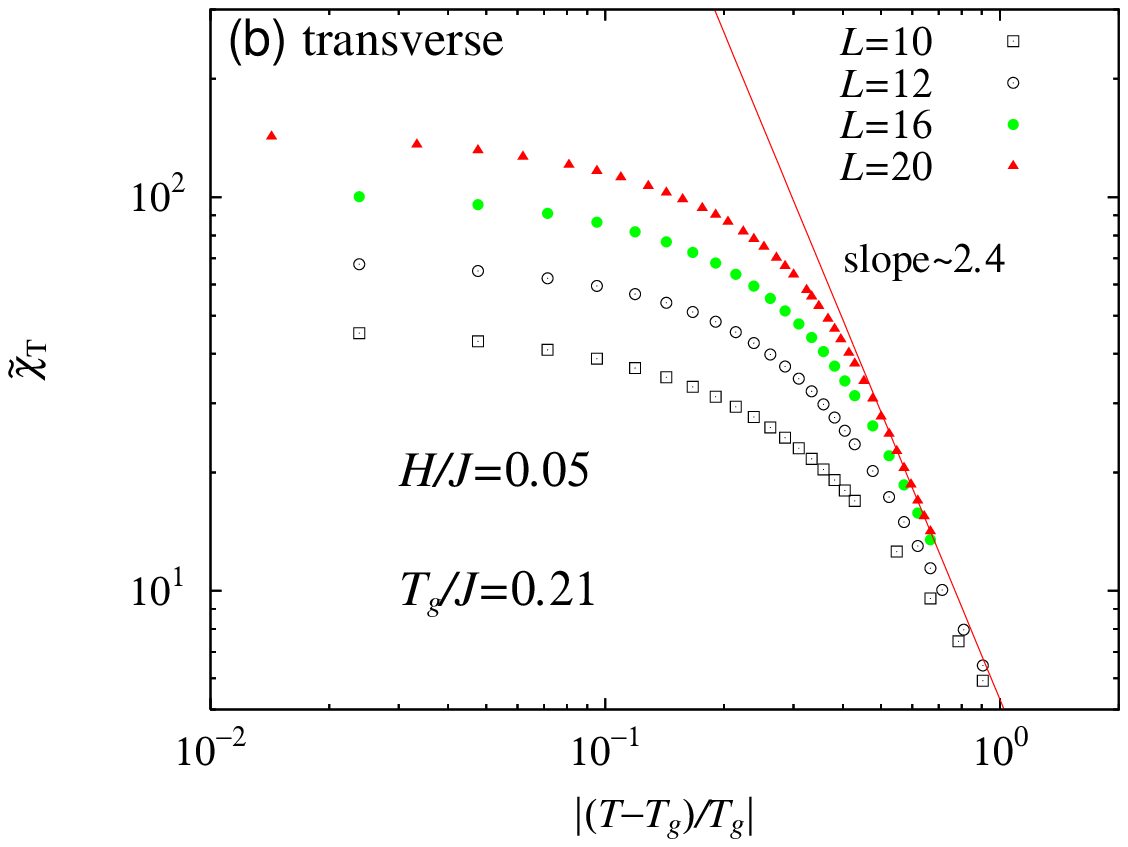}
     \end{tabular}
     \caption{(Color online) 
Temperature and size dependence of (a) the chiral-glass
susceptibility,
     and of (b) the transverse spin-glass
susceptibility, for a field $H/J=0.05$.
The transition temperatures is set to be
$T_g/J=0.21$.}
     \label{fig_XCGSG_D005}
   \end{center}
\end{figure}

In order to estimate the chiral-glass and the spin-glass 
susceptibility exponents
$\gamma_\mathrm{CG}$ and $\gamma_\mathrm{SG}$, 
we plot in Figs.\ref{fig_XCGSG_D005}
(a) the  reduced chiral-glass susceptibility $\tilde \chi _{\chi}$, 
and (b) the transverse spin-glass susceptibility $\tilde \chi_\mathrm{T}$, 
as a function of the reduced
temperature $(T-T_g)/T_g$ for the field $H/J=0.05$ on a log-log plot.
The chiral-glass and the spin-glass susceptibilities 
in the thermodynamic limit are expected 
to behave as
\begin{equation}
\tilde \chi_\chi\approx |T-T_g|^{-\gamma_\mathrm{CG}},\ \ \ \ \ \ 
\tilde \chi_\mathrm{T}\approx |T-T_g|^{-\gamma_\mathrm{SG}}. 
\end{equation}
The $T_g$ value is fixed to be $T_g/J=0.21$. 
From the asymptotic slope of the data, the  exponents $\gamma_\mathrm{CG}$ and
$\gamma_\mathrm{SG}$ are
estimated to be
$\gamma_\mathrm{CG}=2.1(2)$ and
$\gamma_\mathrm{SG}=2.4(2)$, respectively. Here, the obtained values of
$\gamma_\mathrm{SG}$ and $\gamma_\mathrm{CG}$ turn out to be 
rather close. Indeed, 
the spin-chirality decoupling-recoupling scenario predicts an equality 
$\gamma_\mathrm{SG}=\gamma_\mathrm{CG}$.

\subsection{$H$-$T$ phase diagram}

\begin{figure}[ht]
   \begin{center}
     \includegraphics{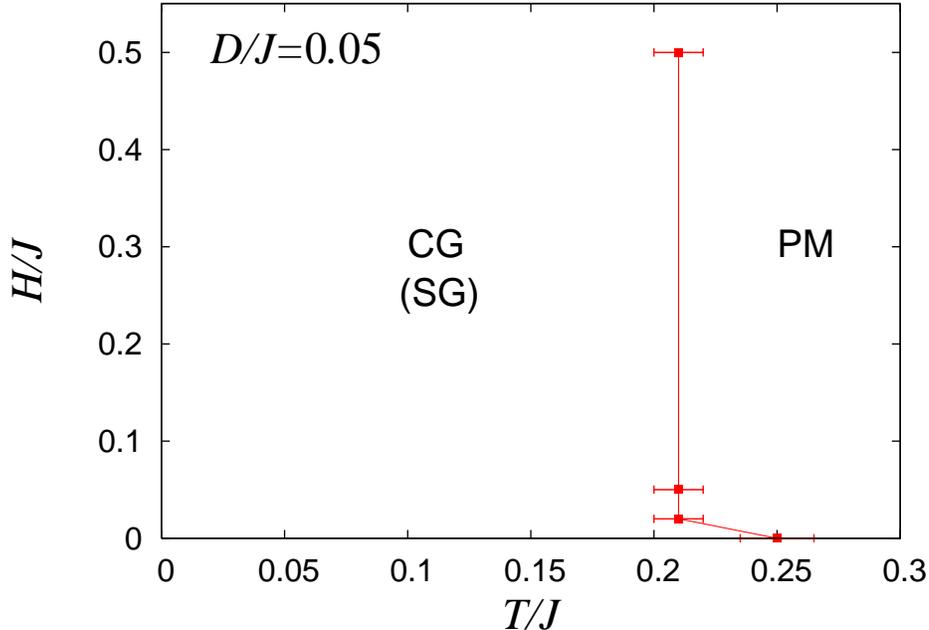}
     \caption{(Color online) 
The phase diagram
of the anisotropic Heisenberg SG with the random anisotropy of $D/J=0.05$ in the temperature-magnetic field plane.
     ``CG'', ``SG'' and ``PM'' stand for the ``chiral-glass'',
     ``spin-glass'' and ``paramagnetic'' phase, respectively.
     }
     \label{fig_HT_phase}
   \end{center}
\end{figure}

Finally,
we show in Fig.\ref{fig_HT_phase}
a phase diagram  of the present model with $D/J=0.05$
in the temperature-magnetic field plane. 
The obtained phase diagram has two noticeable features.
First, for weaker fields $H/J\lsim 0.05=D/J$, the transition line $T_g(H)$ 
exhibits an AT-like like behavior,
{\it i.e.\/}, the spin-glass (chiral-glass) transition temperature
is suppressed rapidly under weak applied fields. In sharp contrast to this, for
stronger fields of $H/J\gsim 0.05=D/J$, the spin-glass (chiral-glass) transition temperature turns out to be rather insensitive to the field intensity, 
indicating
that the spin-glass (chiral-glass) ordered state is
robust against applied fields. The estimated $T_g(H)$ value stays almost unchanged between the field values $H/J=0.05$ and $0.5$, the latter value is
already more than twice the zero-field transition temperature.
Indeed, the $T_g(H)$ value in this field range is very close to the zero-field
chiral-glass transition temperature of the fully {\it isotropic\/} model.
This is exactly the feature of the GT line.

Overall, the obtained
phase diagram is
consistent with the experimental phase diagram of Heisenberg-like SG magnets.
\cite{SGrev,Campbell99,Campbell02}
In order to make a more quantitative comparison, 
however, we have to determine the
phase boundary in the lower field regime $H/J\leq 0.05$ more precisely, and to
map out a phase diagram with varying the anisotropy strength $D$.
This is a computationally demanding task, and is left
as a future task.

\section{Summary and discussion}

In summary,
we performed a large-scale equilibrium Monte Carlo simulation
of the  weakly anisotropic 3D Heisenberg SG in magnetic fields,
paying attention both to the spin and to the chirality. The model is 
expected to be a reasonably realistic model of many real Heisenberg-like
SG magnets. Due to the presence of both the random anisotropy and the magnetic 
field, the model lacks in 
any global symmetry so that the transition, if any, 
is expected to be a pure RSB one.
Among other things, we have found a clear numerical evidence that such a 
pure RSB transition indeed occurs 
at a finite temperature  simultaneously in the spin and in the chiral
sectors. 
The temporal decay of the time correlation functions of the chirality and 
of the spin, as well as the 
overlap distribution function of the chirality and the spin, gave 
a strong numerical evidence of the occurrence of a thermodynamic
RSB transition in applied fields.
We feel this finding for the weakly anisotropic Heisenberg SG
to be rather remarkable, in view of the 
long controversy in the community concerning the in-field ordering
properties of the 
Ising SG, which apparently looks much simpler.

Indeed, a comparison with the 3D Ising SG in fields have revealed that 
the weakly anisotropic
Heisenberg SG in fields behaves  differently from the Ising SG in fields,
a sign of the RSB transition being much clearer in the weakly anisotropic
Heisenberg SG than in the Ising SG.
The observed
stronger sign of the RSB transition is closely connected to the
one-step-like nature of its RSB pattern, which is certainly not the case
in the Ising SG.
Such a difference in the ordering properties of the
two models might at first sound
surprising, since the Ising SG in fields and the weakly anisotropic Heisenberg 
SG in fields share the same symmetry property. 
However, 
in spite of the similarity in symmetry, a significant difference exists in the
two models in 
that the Ising SG does not possess
any chirality degree of freedom in contrast to the 
weakly anisotropic Heisenberg SG. Then, our observation 
highlights the possible importance of 
the chirality  in realizing the RSB transition in the
weakly anisotropic Heisenberg SG. 
This gives a strong support to the chirality scenario of the SG transition
of Heisenberg-like SGs.\cite{Kawamura92,Kawamura98}

We also have found that the sign of the RSB transition
is often observed  more strongly in the chirality-related quantities 
than in the spin-related quantities.
For example, the second peak of the overlap distribution function shows up 
clearly even from smaller lattices for the chirality, 
but more weakly and only for larger lattices
for the spin: Compare Fig.8(a) and Fig.9(a). This observation
seems consistent with the chirality scenario that the order parameter of 
the transition is the chirality whereas  
the spin is recoupled to the chirality
beyond a certain crossover (recoupling) length scale of 
about $20$ lattice spacings.
We have also observed that
the behaviors of certain 
spin-related quantities, {\it e.g.\/}, the
Binder ratio or the normalized correlation length, 
differ between in smaller and in 
larger lattices, the borderline size being around $L\simeq 20$.
Again, these behaviors 
could naturally be explicable on the basis of
the spin-chirality decoupling-recoupling scenario.\cite{Kawamura92,Kawamura98}

We also have constructed a rough
$H$-$T$ phase diagram of the model. It has turned out that
$T_g(H)$ is rapidly suppressed 
by weak applied fields, $H\lsim D$, but  is then kept almost 
unchanged up to rather higher fields of $H/J=0.5\approx 2T_g(0)$. 
The phase boundary in the
weak field regime resembles the AT-line, while that in the high field regime
resembles the GT line, although its nature is very different from the
mean-field AT- or GT-line.\cite{KotSom,AT,GT} 
These features of the phase diagram
are consistent with the experimental result of the weakly anisotropic
Heisenberg SG. \cite{Campbell99,Campbell02}
The obtained phase diagram is  also fully consistent with the
one expected from the chirality scenario. 
\cite{Kawamura92,Kawamura98,KawaIma01,ImaKawa04}
Indeed,
the chirality scenario predicts that, in the low-field regime where
the anisotropy is important, the change in the broken symmetry from the
zero-field case causes a crossover, accompanied with a rapid suppression of the
transition temperature, while in the high-field regime where the
anisotropy is negligible relative to the applied magnetic field,
the SG transition line should essentially
be given by the chiral-glass transition
line of the fully isotropic system.\cite{KawaIma01}

Another remarkable feature of the present phase diagram is that the
GT-like phase boundary is quite robust against magnetic fields, 
extending without much reduction to the field at least as twice 
large as the zero-field transition temperature. Such a robustness
of the SG ordered phase under fields is also in accord with the chirality
scenario, since the chirality, regarded as a hidden order parameter of the 
SG transition, is only weakly coupled to the magnetic field.
Note that the magnetic field couples directly to the 
spin via the Zeeman term in the Hamiltonian, 
not to the chirality, while the effective
coupling between the field and the chirality is rather weak.

Thus, overall, our observations on the weakly anisotropic Heisenberg
SG gives a strong support to the spin-chirality decoupling-recoupling
scenario. \cite{Kawamura92,Kawamura98} In other words, it would be difficult
to interpret all of our present
observations based on the more conventional view that the spin is 
the order parameter of the transition and the chirality is just a
composite (secondary). 

A decisive test of the 
spin-chirality decoupling-recouping scenario would be to directly probe
the chiral-glass order.
Recently, a proposal was made concerning the way to  measure the 
linear and the nonlinear chiral susceptibilities of canonical SGs 
by using the Hall probe, 
together with the
scaling prediction from the spin-chirality decoupling-recoupling 
scenario. \cite{Kawamura03,Tatara02}  Interestingly,
very recent experiments have given support to these predictions from
the chirality scenario. \cite{Taniguchi03,Campbell03,Sato}

In concluding this paper, we wish to 
emphasize again that, even if one sets aside the question of 
the detailed 
mechanism of the SG transition or the validity of the chirality scenario, 
the present numerical results have
demonstrated beyond reasonable doubt 
that the weakly anisotropic Heisenberg SG exhibits
a thermodynamic RSB transition in applied magnetic fields.
In view of the fact that the property
of the apparently simpler, strongly anisotropic Ising SG 
still remains unclear in spite of the long controversy,
this finding seems surprising and somewhat ironical. 
Clarifying 
the relation between the weakly anisotropic Heisenberg SG possessing
the chiral degree of freedom and the strongly
anisotropic Ising SG not possessing the chiral degree of freedom
is still open, and is left for future studies.

\section*{Acknowledgments}

The numerical calculation was performed on the HITACHI
SR8000 at the supercomputer system, ISSP, University of Tokyo,
and Pentium IV clustering machines in our laboratory.
The authors are thankful to I.A. Campbell, K. Hukushima, H. Yoshino, 
and G. Tatara for useful discussion and comments.



\end{document}